\documentclass[twocolumn]{aastex63}

\received{\ldots}
\revised{\ldots}
\accepted{\ldots}
\published{\ldots}
\reportnum{Accepted to ApJL}

\shorttitle{The X-ray reactivation of \srclong}
\shortauthors{A. Borghese et al.}

\begin{document}

\title{THE X-RAY REACTIVATION OF THE RADIO BURSTING MAGNETAR \srclong}



\correspondingauthor{A. Borghese}
\email{borghese@ice.csic.es}
\author[0000-0001-8785-5922]{A. Borghese}
\affil{Institute of Space Sciences (ICE, CSIC), Campus UAB, Carrer de Can Magrans s/n, E-08193, Barcelona, Spain}
\affil{Institut d'Estudis Espacials de Catalunya (IEEC), Carrer Gran Capit\`a 2--4, E-08034 Barcelona, Spain}

\author[0000-0001-7611-1581]{F. Coti Zelati}
\affil{Institute of Space Sciences (ICE, CSIC), Campus UAB, Carrer de Can Magrans s/n, E-08193, Barcelona, Spain}
\affil{Institut d'Estudis Espacials de Catalunya (IEEC), Carrer Gran Capit\`a 2--4, E-08034 Barcelona, Spain}

\author[0000-0003-2177-6388]{N. Rea}
\affil{Institute of Space Sciences (ICE, CSIC), Campus UAB, Carrer de Can Magrans s/n, E-08193, Barcelona, Spain}
\affil{Institut d'Estudis Espacials de Catalunya (IEEC), Carrer Gran Capit\`a 2--4, E-08034 Barcelona, Spain}

\author[0000-0003-4849-5092]{P. Esposito}
\affil{Scuola Universitaria Superiore IUSS Pavia, Palazzo del Broletto, piazza della Vittoria 15, I-27100 Pavia, Italy}
\affil{INAF--Istituto di Astrofisica Spaziale e Fisica Cosmica di Milano, via A.\,Corti 12, I-20133 Milano, Italy}

\author[0000-0001-5480-6438]{G. L. Israel}
\affil{INAF--Osservatorio Astronomico di Roma, via Frascati 33, I-00078 Monteporzio Catone, Italy}

\author[0000-0003-3259-7801]{S. Mereghetti}
\affil{INAF--Istituto di Astrofisica Spaziale e Fisica Cosmica di Milano, via A.\,Corti 12, I-20133 Milano, Italy}

\author[0000-0002-6038-1090]{A. Tiengo}
\affil{Scuola Universitaria Superiore IUSS Pavia, Palazzo del Broletto, piazza della Vittoria 15, I-27100 Pavia, Italy}
\affil{INAF--Istituto di Astrofisica Spaziale e Fisica Cosmica di Milano, via A.\,Corti 12, I-20133 Milano, Italy}
\affil{Istituto Nazionale di Fisica Nucleare (INFN), Sezione di Pavia, via A.\,Bassi 6, I-27100 Pavia, Italy}

\def\xmm {\emph{XMM--Newton}}
\def\cxo {\emph{Chandra}}
\def\nustar {\emph{NuSTAR}}
\def\rst {\emph{ROSAT}}
\def\rxte {\emph{RXTE}}
\def\swift {\emph{Swift}}
\def\nicer {\emph{NICER}}

\def\srclong{SGR\,J1935+2154}
\def\src{SGR\,J1935}

\def\flux {\mbox{erg cm$^{-2}$ s$^{-1}$}}
\def\lum {\mbox{erg s$^{-1}$}}
\def\nh {$N_{\rm H}$}
\def\kms  {\rm \ km \, s^{-1}}
\def\cms  {\rm \ cm \, s^{-1}}
\def\gs   {\rm \ g  \, s^{-1}}
\def\cmtre {\rm \ cm^{-3}}
\def\cmdue {\rm \ cm$^{-2}$}
\def\ss {\mbox{s\,s$^{-1}$}}
\def\chisq {$\chi ^{2}$}
\def\rchisq {$\chi_{\nu} ^{2}$}

\def\arc{\mbox{$^{\prime\prime}$}}
\def\arcmin{\mbox{$^{\prime}$}}
\def\deg{\mbox{$^{\circ}$}}

\def\rsun {~R_{\odot}}
\def\msun {~M_{\odot}}
\def\mdotav {\langle \dot {M}\rangle }

\def\uu {4U\,0142$+$614}
\def\ee {1E\,1048.1$-$5937}
\def\kes {1E\,1841$-$045}
\def\aa {1E\,1547$-$5408}
\def\axj {AX\,J1844$-$0258}
\def\rxs {1RXS\,J1708$-$4009}
\def\xte{XTE\,J1810$-$197}
\def\smc{CXOU\,J0100$-$7211}
\def\wes{CXOU\,J1647$-$4552}
\def\ea {1E\,2259$+$586}
\def\sgra{SGR\,1806$-$20}
\def\sgrb{SGR\,1900$+$14}
\def\sgrd{SGR\,1627$-$41}
\def\sgre{SGR\,0501$+$4516}
\def\lowba{SGR\,0418$+$5729}
\def\sgrg{SGR\,1833$-$0832}
\def\lowbb{Swift\,J1822.3$-$1606}
\def\galmag{PSR\,J1745$-$2900}
\def\sgras{Sgr\,A$^{\star}$}
\def\sgrh{SGR\,1801$-$21}
\def\sgri{SGR\,2013$+$34}
\def\psr{PSR\,J1622$-$4950}
\def\hbpsr{PSR\,J1846$-$0258}
\def\radiohb{PSR\,J1119$-$6127}
\def\coronamag{Swift\,J1818.0$-$1607}

\newcommand{\com}[1]{\textcolor{red}{#1}}

\begin{abstract}
A few years after its discovery as a magnetar, \srclong\ started a new burst-active phase on 2020 April 27, accompanied by a large enhancement of its X-ray persistent emission. Radio single bursts were detected during this activation, strengthening the connection between magnetars and fast radio bursts. We report on the X-ray monitoring of \srclong\ from $\sim$3 days prior to $\sim$3 weeks after its reactivation, using \swift, \nustar, and \nicer. We detected X-ray pulsations in the \nicer\ and \nustar\ observations, and constrained the spin period derivative to $|\dot{P}| < 3 \times 10^{-11}$\,s\,s$^{-1}$ (3$\sigma$ c.l.). The pulse profile showed a variable shape switching between single and double-peaked as a function of time and energy. The pulsed fraction decreased from $\sim$34\% to $\sim$11\% (5--10\,keV) over $\sim$10 days. The X-ray spectrum was well fit by an absorbed blackbody model with temperature decreasing from $kT_{\rm BB} \sim$ 1.6 to 0.45--0.6\,keV, plus a nonthermal power-law component ($\Gamma\sim 1.2$) observed up to $\sim$25\,keV with \nustar. The 0.3--10\,keV X-ray luminosity increased in less than four days from $\sim 6\times 10^{33} d^2_{6.6}$\,\lum\ to about 3$\times10^{35} d^2_{6.6}$\,\lum\ and then decreased again to 2.5$\times 10^{34} d^2_{6.6}$\,\lum\ over the following three weeks of the outburst, where $d_{6.6}$ is the source distance in units of 6.6\,kpc. We also detected several X-ray bursts, with properties typical of short magnetar bursts. 
\end{abstract}


\section{Introduction} \label{sec:intro}

\begin{deluxetable*}{ccccccccc}
\tablecaption{Observation log and blackbody spectral parameters. 
\label{tab:observations}}
\tabletypesize{\scriptsize}
\tablenum{1}
\tablewidth{0pt}
\tablehead{
\colhead{Instrument\tablenotemark{\footnotesize a}} &
\colhead{Obs.ID} &
\colhead{Start} &
\colhead{Stop} & 
\colhead{Exposure} & 
\colhead{Count Rate\tablenotemark{\footnotesize b}} &
\colhead{$kT_{\rm BB}$} &
\colhead{$R_{\rm BB}$} &
\colhead{Flux\tablenotemark{\footnotesize c}} \\
\colhead{} &
\colhead{} & 
\multicolumn{2}{c}{YYYY-MM-DD hh:mm:ss (TT)} &
\colhead{(ks)} & 
\colhead{(counts\,s$^{-1}$)} &
\colhead{(keV)} &
\colhead{(km)} &
\colhead{(10$^{-11}$\,cgs)}
}
\startdata
\swift/XRT (PC) & 00033349044 & 2020-04-23 15:16:16 & 2020-04-23 15:49:27 & 2.0 & 0.012$\pm$0.002 & -- & -- & 0.045\tablenotemark{\footnotesize d} \\
\swift/XRT (PC) & 00968211001 & 2020-04-27 19:41:56 & 2020-04-27 20:15:09 & 1.8 & 0.37$\pm$0.01 & 1.6$_{-0.1}^{+0.2}$  & 0.49$_{-0.10}^{+0.12}$ & 5.01$_{-0.59}^{+0.05}$ \\
\nicer/XTI      & 3020560101  & 2020-04-28 00:38:31 & 2020-04-28 16:21:20 & 4.7 & 2.94$\pm$0.04 & 0.80$\pm$0.02 & 1.11$\pm$0.05 & 2.49$\pm$0.12\\
\swift/XRT (PC) & 00033349045 & 2020-04-28 18:00:36 & 2020-04-28 21:37:41 & 2.9 & 0.077$\pm$0.005 & 0.61$_{-0.10}^{+0.09}$ & 0.99$_{-0.19}^{+0.39}$ & 0.6$\pm$0.1 \\
\swift/XRT (WT) & 00033349046 & 2020-04-29 13:07:57 & 2020-04-29 13:32:57 & 1.5 & 0.09$\pm$0.01 & 0.36$_{-0.10}^{+0.15}$ & 2.74$_{-1.08}^{+6.61}$ & 0.69$_{-0.15}^{+0.09}$ \\
\nicer/XTI      & 3020560102  & 2020-04-29 13:47:17 & 2020-04-29 14:05:20 & 1.1 & 0.96$\pm$0.05 & 0.49$\pm$0.02\tablenotemark{\footnotesize e}  & 1.47$_{-0.14}^{+0.17}$ & 0.82$_{-0.08}^{+0.11}$ \\
\swift/XRT (PC) & 00033349047 & 2020-04-29 17:54:22 & 2020-04-29 18:27:38 & 2.0 & 0.072$\pm$0.006 & 0.44$_{-0.06}^{+ 0.07}$ & 2.05$_{-0.49}^{+1.04}$ & 0.52$_{-0.10}^{+0.08}$ \\
\nicer/XTI      & 3655010101 & 2020-04-29 21:31:57 & 2020-04-29 21:48:40 & 0.8 & 0.78$\pm$0.04 & 0.49$\pm$0.02\tablenotemark{\footnotesize e}   & 1.88$_{-0.11}^{+0.15}$ & 0.40$\pm$0.07 \\
\nicer/XTI      & 3655010102 & 2020-04-30 00:37:56 & 2020-04-30 07:09:40 & 5.3 & 0.73$\pm$0.02 & 0.49$\pm$0.02\tablenotemark{\footnotesize e}   & 1.73$_{-0.09}^{+0.12}$ & 0.40$_{-0.07}^{+0.05}$ \\
\nicer/XTI      & 3020560103 & 2020-04-30 13:02:45 & 2020-04-30 13:17:20 & 0.8 & 0.72$\pm$0.04 & 0.49$\pm$0.02\tablenotemark{\footnotesize e}    & 1.83$_{-0.14}^{+0.16}$ & 0.32$_{-0.07}^{+0.13}$  \\
\swift/XRT (PC) & 00033349048 & 2020-04-30 05:29:05  & 2020-04-30 18:27:53 & 1.9 & 0.054$\pm$0.005 & 0.44$_{-0.09}^{+0.11}$ & 1.55$_{-0.48}^{+1.56}$ & 0.46$_{-0.13}^{+0.02}$ \\
\swift/XRT (WT) & 00033349049 & 2020-04-30 07:10:24 & 2020-04-30 11:47:56 & 1.5 & 0.05$\pm$0.01 & 0.40$_{-0.40}^{+0.27}$ & 1.84$_{-0.83}^{+8.70}$ & 0.57$\pm$0.16 \\
\swift/XRT (PC) & 00033349050 & 2020-05-01 02:03:14 & 2020-05-01 22:42:20 & 2.1 & 0.056$\pm$0.005 & 0.63$_{-0.16}^{+0.09}$ & 0.93$_{-0.19}^{+0.51}$ & 0.36$_{-0.09}^{+0.19}$ \\
\swift/XRT (WT) & 00033349051 & 2020-05-01 12:58:08 & 2020-05-01 13:20:56 & 1.4 & 0.05$\pm$0.01 & 0.39$_{-0.08}^{+0.09}$ & 2.98$_{-0.88}^{+2.74}$ & 0.33$_{-0.15}^{+0.04}$ \\
\nustar\ FPMA/B & 80602313002  & 2020-05-02 00:06:09 & 2020-05-02 20:31:09 & 37.1/36.9 & 0.175$\pm$0.003 & 0.59$_{-0.05}^{+0.06}$ & 0.85$_{-0.18}^{+0.35}$ & 0.32$\pm$0.01 \\ 
\swift/XRT (WT) & 00033349053 & 2020-05-02 11:50:05 & 2020-05-02 13:28:56 & 0.7 & 0.06$\pm$0.02 & 0.69$_{-0.18}^{+0.14}$ & 0.97$_{-0.26}^{+0.52}$ & 0.30$_{-0.09}^{+0.52}$ \\
\swift/XRT (PC) & 00033349052 & 2020-05-02 16:33:41 & 2020-05-02 23:02:54 & 1.2 & 0.027$\pm$0.005 & 0.71$_{-0.27}^{+0.14}$ & 0.69$_{-0.22}^{+0.32}$ & 0.18$_{-0.01}^{+0.63}$ \\
\swift/XRT (WT) & 00033349055 & 2020-05-03 12:55:54 & 2020-05-03 13:23:56 & 1.7 & 0.020$\pm$0.009 & 0.45$_{-0.15}^{+0.23}$ & 1.24$_{-0.50}^{+3.10}$ & 0.27$_{-0.15}^{+0.06}$ \\
\swift/XRT (PC) & 00033349054 & 2020-05-03 22:23:05 & 2020-05-03 22:48:52 & 1.5 & 0.050$\pm$0.006 & 0.68$_{-0.18}^{+0.02}$ & 0.93$_{-0.18}^{+0.41}$ & 0.24$_{-0.01}^{+0.4}$ \\ 
\swift/XRT (PC) & 00033349056 & 2020-05-04 01:47:23 & 2020-05-04 18:04:51 & 3.4 & 0.040$\pm$0.003 & 0.48$_{-0.07}^{+0.08}$ & 1.29$_{-0.29}^{+0.65}$ & 0.28$_{-0.06}^{+0.03}$ \\
\swift/XRT (WT) & 00033349057 & 2020-05-04 12:40:56 & 2020-05-04 13:07:56 & 1.6 & 0.07$\pm$0.01 & 0.66$_{-0.20}^{+0.18}$ & 0.85$_{-0.24}^{+0.97}$ & 0.53$_{-0.20}^{+0.05}$ \\
\swift/XRT (PC) & 00033349058 & 2020-05-05 03:17:19 & 2020-05-05 13:01:52 & 1.9 & 0.034$\pm$0.004 & 0.53$_{-0.17}^{+0.09}$ & 1.15$_{-0.27}^{+1.82}$ & 0.15$_{-0.09}^{+0.15}$ \\
\swift/XRT (WT) & 00033349059 & 2020-05-05 20:40:09 & 2020-05-05 21:02:56 & 1.4  & 0.05$\pm$0.01 & 0.55$_{-0.07}^{+0.08}$ & 1.37$_{-0.31}^{+0.54}$ & 0.19$_{-0.04}^{+0.07}$  \\
\swift/XRT (PC) & 00033349060 & 2020-05-06 06:36:44 & 2020-05-06 08:20:52 & 1.3 & 0.031$\pm$0.005 & 0.46$_{-0.08}^{+0.10}$ & 1.50$_{-0.42}^{+1.06}$ & 0.18$_{-0.07}^{+0.04}$ \\
\swift/XRT (PC) & 00033349061 & 2020-05-07 09:30:09 & 2020-05-07 20:56:54 & 3.7 & 0.035$\pm$0.003 & 0.47$\pm$0.09 & 1.57$_{-0.33}^{+0.99}$ & 0.23$_{-0.13}^{+0.08}$    \\
\swift/XRT (PC) & 00033349062 & 2020-05-10 04:28:08 & 2020-05-10 22:15:52 & 3.2 & 0.043$\pm$0.004 & 0.63$_{-0.13}^{+0.02}$ & 0.99$_{-0.16}^{+0.32}$ & 0.23$\pm$0.11 \\
\swift/XRT (WT) & 00033349063 & 2020-05-10 06:01:43 & 2020-05-10 10:56:56 & 3.2 & 0.030$\pm$0.007 & 0.51$\pm$0.10 & 1.51$_{-0.37}^{+1.06}$ & 0.24$_{-0.08}^{+0.05}$    \\
\nustar\ FPMA/B  & 80602313004 & 2020-05-10 23:51:09 & 2020-05-11 20:31:09 & 38.5/38.2 & 0.140$\pm$0.002 & 0.52$\pm$0.04 & 1.03$_{-0.20}^{+0.32}$ &  0.27$\pm$0.01 \\ 
\nicer/XTI      & 3020560104 & 2020-05-11 14:30:54 & 2020-05-11 16:18:40 & 1.3 & 0.54$\pm$0.04 & 0.49$\pm$0.02\tablenotemark{\footnotesize e}  & 1.70$_{-0.11}^{+0.14}$ & 0.16$\pm$0.01  \\
\swift/XRT (WT) & 00033349064 & 2020-05-13 02:22:52 & 2020-05-13 07:29:55 & 1.9  & 0.56$_{-0.17}^{+0.14}$\tablenotemark{\footnotesize f} & 0.69$\pm$0.10 & 0.95$_{-0.26}^{+1.29}$ & 0.17$_{-0.06}^{+0.03}$\\
\swift/XRT (WT) & 00033349065 & 2020-05-13 09:03:52 & 2020-05-13 10:30:56 & 1.3  & 0.56$_{-0.17}^{+0.14}$\tablenotemark{\footnotesize f} & 0.69$\pm$0.10 & 0.95$_{-0.26}^{+1.29}$ & 0.17$_{-0.06}^{+0.03}$ \\
\swift/XRT (WT) & 00033349066 & 2020-05-15 00:31:07 & 2020-05-15 03:58:39 & 3.5 & 0.059$\pm$0.006 & 0.46$\pm$0.07 & 1.70$_{-0.39}^{+0.86}$ &  0.24$_{-0.10}^{+0.04}$   \\
\enddata
\tablenotetext{a}{The instrumental setup is indicated in brackets: PC = photon counting, WT = windowed timing.}
\tablenotetext{b}{Count rate, computed after removing bursts, in the 0.3--10\,keV range for \swift, in the 1--5\,keV band for \nicer, and in the 3--25\,keV range for \nustar\ summing up the two FPMs.}
\tablenotetext{c}{Observed 0.3--10\,keV flux in units of 10$^{-11}$\,\flux.}
\tablenotetext{d}{The flux is estimated using {\sc webpimms} (see text for details).}
\tablenotetext{e}{The blackbody temperature was tied up among these data sets (see text for details).}
\tablenotetext{f}{These observations were combined to increase the signal-to-noise.}
\end{deluxetable*}

Magnetars are isolated X-ray pulsars with spin periods in the 0.3--12\,s range and large spin-down rates, implying particularly strong surface dipolar magnetic fields of the order of $B\sim10^{14}$--10$^{15}$\,G \citep[see][for recent reviews]{kaspi17,esposito18}. These objects have a persistent X-ray luminosity of $L_{\mathrm{X}}\sim 10^{31}$--10$^{36}$\,\lum, which is thought to be powered by the instabilities and decay of their extreme magnetic fields. Among isolated neutron stars, magnetars are the most variable, with an unpredictable bursting activity. They emit short ($<$1\,s) and bright ($L_{{\rm peak}}$ $\approx$ 10$^{39}$--10$^{41}$\,\lum) bursts in the X-ray band, either sporadically or clustered in ``forests'' \citep[e.g.,][]{israel08,collazzi15}. These bursts are often accompanied by an enhancement of the X-ray persistent flux, up to three orders of magnitude above quiescence. Then, the flux usually relaxes back to the pre-outburst level on months/years timescales \citep{cotizelati18}. Recently, magnetar traits have been observed also in high-$B$ pulsars \citep[e.g.,][]{gavriil08,archibald16}, X-ray pulsars with dipolar fields as low as $6\times10^{12}$\,G \citep[e.g.,][]{rea10,rea12a}, and the central source of the supernova remnant RCW\,103 \citep[e.g.,][]{rea16,dai16,borghese18}. These findings have shown how magnetar-like emission might be more common within the neutron star population than previously expected.     

\srclong\ (\src\ hereafter) was discovered in 2014, when the Burst Alert Telescope (BAT) on board of the \emph{Neil Gehrels Swift Observatory} \citep{gehrels04} triggered on a short burst \citep{stamatikos14}. A follow-up campaign confirmed the source as a magnetar with spin period $P\sim3.25$\,s and spin-down rate $\dot{P}\sim1.43 \times 10^{-11}$\,\ss, implying a dipole magnetic field $B\sim4.4 \times 10^{14}$\,G at the pole and characteristic age $\tau_{\mathrm{c}}\sim3.6$\,kyr \citep{israel16}. \src\ has been quite active since then, with intense outbursts in February 2015, May and June 2016 \citep{younes17}, and frequent bursting activity \citep{lin20a}. 

\src\ reactivated on 2020 April 27--28, emitting a forest of X-ray bursts \citep[e.g.,][]{palmer20,younes20} accompanied by an increase of the persistent X-ray flux, as typical in magnetar outbursts. More interestingly, two millisecond radio bursts temporally coincident with a hard X-ray burst were detected from the direction of the source \citep{scholz20,bochenek20,li20,mereghetti20,tavani20}, strenghtening the long suspected connection between magnetars and fast radio bursts (FRBs; see \citealt{cordes19,petroff19} for reviews). However, besides these radio bursts, radio pulsed emission has not been detected so far from the source \citep[e.g.,][]{younes17,lin20b}.

This Letter reports on the results of our monitoring campaign of \src\ with \swift, \nustar, and \nicer, covering the first $\sim$20 days since its reactivation. We describe the observations (\S\ref{sec:obs}) and report our timing and spectral analysis as well as a search for short bursts (\S\ref{sec:analysis}). We discuss our findings in \S\ref{sec:discuss}.

\begin{figure*}
    \centering
    \includegraphics[width=2.1\columnwidth]{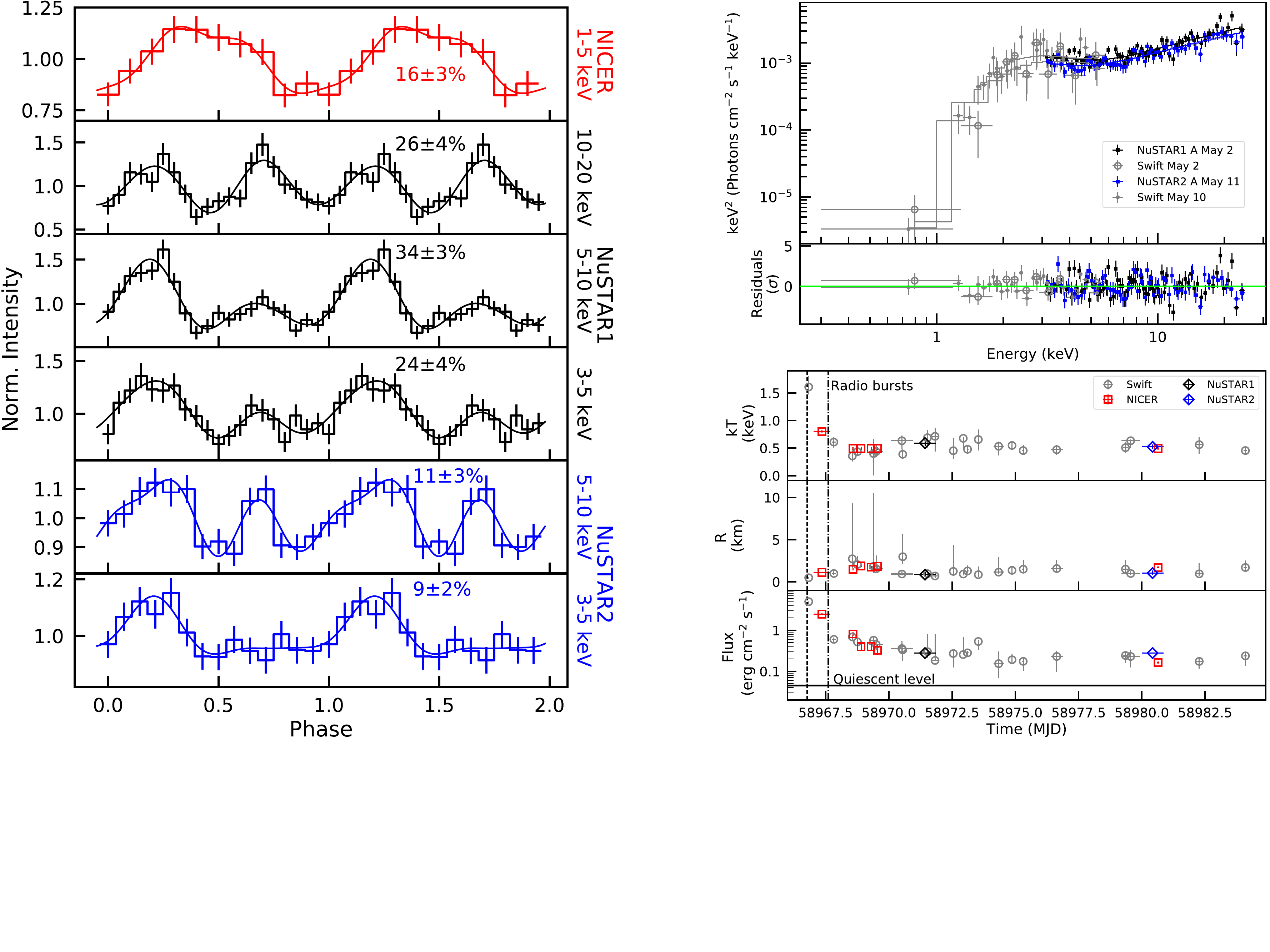}
    \vspace{-3cm}
    \caption{{\em Left:} Energy-resolved background-subtracted pulse profiles of \src\ extracted from \nicer\ and \nustar\ data. The profiles at the different epochs have been aligned so as to have the pulse minimum at phase 0. The best-fitting models obtained by using two (for \nicer) and three (\nustar) sinusoidal components (fundamental plus harmonics) are shown with solid lines. The corresponding pulsed fractions are reported in each panel. {\em Right-top:} Broad-band unfolded spectra extracted from the quasi-simultaneous \swift/XRT and \nustar\ data on 2020 May 2 and 11. The best-fitting model is plotted with a solid line. We show only the FPMA spectra for displaying purpose. The {\em bottom} panel shows the post-fit residuals in units of standard deviations. {\em Right-bottom}: Temporal evolution of the blackbody temperature ({\em top}), radius ({\em middle}), and observed flux in units of 10$^{-11}$\,\flux\ (0.3--10\,keV; {\em bottom}). The dashed line denotes the epoch of the first BAT trigger (MJD 58966.7683; \citealt{palmer20}). The dashed-dotted line marks the epoch of the two bright radio bursts (MJD 58967.6072; \citealt{scholz20,bochenek20}). The solid line in the {\em bottom} panel marks the quiescent flux, $\sim 4.5 \times 10^{-13}$\,\flux.}
    \label{fig:spec}
\end{figure*}

\section{Observations and data reduction} \label{sec:obs}
We report the log of the observations used in this work in Table\,\ref{tab:observations}. Data reduction was performed using tools in the {\sc heasoft} package (version 6.27.2).
Photon arrival times were referred to the solar system barycenter using the source \cxo\ position (RA = 19$^\mathrm{h}$34$^\mathrm{m}$55$\fs$598, Dec = +21$^{\circ}$53$^{\prime}$47$\farcs$79, J2000.0; \citealt{israel16}) and the JPL planetary ephemeris DE\,200. In the following, we adopt a distance of 6.6\,kpc \citep[][see also \citealt{mereghetti20}]{zhou20} and quote all uncertainties at 1$\sigma$ confidence level (c.l.).

\subsection{\swift} 
After the \swift/BAT trigger, \src\ was monitored almost daily with the \swift/XRT \citep{burrows05} either in photon counting (PC; timing resolution of 2.51\,s) or windowed timing (WT; 1.8\,ms) modes. The data were reprocessed and analysed with standard prescriptions. 

In the first XRT observation performed after the BAT trigger, a dust scattering 
ring was detected around the source, extending from $\sim$1 to 2 arcmin \citep{kennea20,mereghetti20}. This structure was no longer observed in a pointing performed the following day (a detailed study of this structure will be presented in a future paper). We collected the source photons from a 20-pixel circle (1 pixel $ = 2\farcs36$). Background counts were extracted from a region of the same size for WT data and an annulus with radii of 100 and 150 pixels, centered on the source, for the PC observations.

\subsection{\nustar} 

\src\ was observed with \nustar\ \citep{harrison13} twice, on 2020 May 2 and 11. The two focal plane modules FPMA and FPMB observed the source for a total on-source exposure time of 75.6 and 75.1\,ks, respectively. We used the tool {\sc nupipeline} to create cleaned event files and filter out passages through the South Atlantic Anomaly. The source counts were collected within a circular region of radius 100 arcsec, while the background was estimated from a 100-arcsec circle on the same chip of the target. In both pointings, \src\ is detected until $\sim$25\,keV. We ran the script {\sc nuproducts} to extract light curves and spectra, and generate response files for both FPMs.     

\subsection{\nicer}

\nicer\ \citep{gendreau12} observed \src\ six times for a total on-source exposure time of $\sim$ 14\,ks. 
The data were processed via the {\sc nicerdas} pipeline, with the tool {\sc nicerl2} with standard filtering criteria. The background count rate and spectra were computed from \nicer\ observations of the \rxte\ blank-field regions using {\sc nibackgen3C50}.

\section{Analysis and results}\label{sec:analysis}

\subsection{Timing Analysis}\label{sec:timing}
For the timing analysis, we selected events in the 1--5\,keV energy band for \nicer\ and 3--20\,keV for \nustar. 
The data sets of \nicer\ observations IDs. 3655010101, 3655010102 and 3020560103 performed on April 29--30 were merged to increase the source signal-to-noise ratio. We did not include \swift/XRT observations in our timing analysis due to their poor counting statistics.

We calculated a power density spectrum (PDS) for all time series to search for the spin signal, assuming a 3.5$\sigma$ detection threshold for the signal (using the algorithm by \citealt{israel96}), taking into account all the frequencies in the PDS. Pulsations were significantly detected over a blind search only during the first \nustar\ observation. The signal was then found in the second \nustar\ observation and in the \nicer\ combined pointings IDs. 3655010101 $+$ 3655010102 $+$ 3020560103 by looking in the range of periods $P\pm \Delta P$ (at 3$\sigma$; the $\dot{P}$ component can be neglected) around the value measured in the first \nustar\ data set. The period values were then refined by means of a phase-fitting technique. We obtained the following results: $P$ =  3.24731(1)\,s for the combined \nicer\ data sets (April 29--30), $P$ = 3.247331(3)\,s for the first \nustar\ observation (May 2) and $P$ = 3.24734(1)\,s for the second \nustar\ observation (May 11). The above uncertainties and the variable pulse profile (see below) did not allow us to phase-connect coherently the \nicer\ and \nustar\ observations. 
These period measurements imply an upper limit on the spin period derivative of $|\dot{P}| < 3 \times 10^{-11}$\,s\,s$^{-1}$ (3$\sigma$ c.l.), a factor of about two above the value inferred during the 2014 outburst \citep{israel16}. 

Fig.\,\ref{fig:spec} shows the pulse profiles at different epochs and as a function of energy. The profile shape varies considerably in time, changing from quasi-sinusoidal on April 29-30 to double-peaked on May 2 and 11 (the separation between the two peaks is about half rotational cycle). The profile shape is also highly variable with energy in the \nustar\ data sets, the second peak (at phase $\sim$0.6--0.7) being more prominent above 5\,keV and dominating above 10\,keV in the first observation.

The background-subtracted pulsed fraction (defined as the semi-amplitude of the sinusoidal functions describing the pulse divided by the source average count rate) decreased by a factor of $\approx3$ between May 2 and 11 (in the 3--5 and 5--10\,keV ranges; see Fig.\,\ref{fig:spec}). No pulsations were detected over the 10--20\,keV band in the second \nustar\ observation, and we set a 3$\sigma$ upper limit on the pulsed fraction of $\sim$15\%. 

\begin{figure*}
    \centering
    \vspace{-0.3cm}
    \includegraphics[width=1.45\columnwidth]{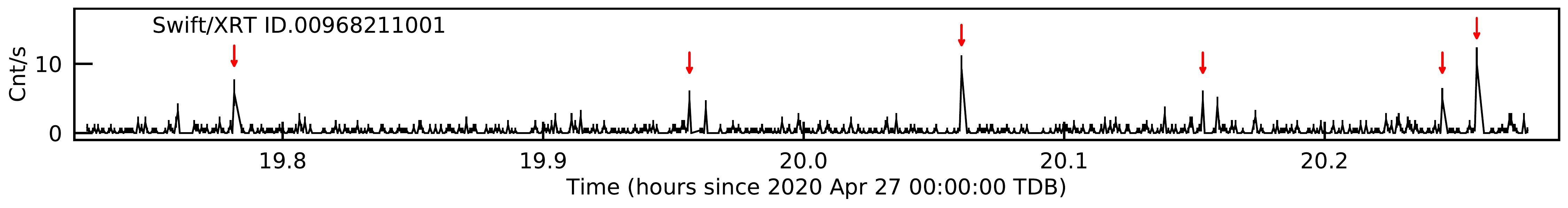}
    \includegraphics[width=1.45\columnwidth]{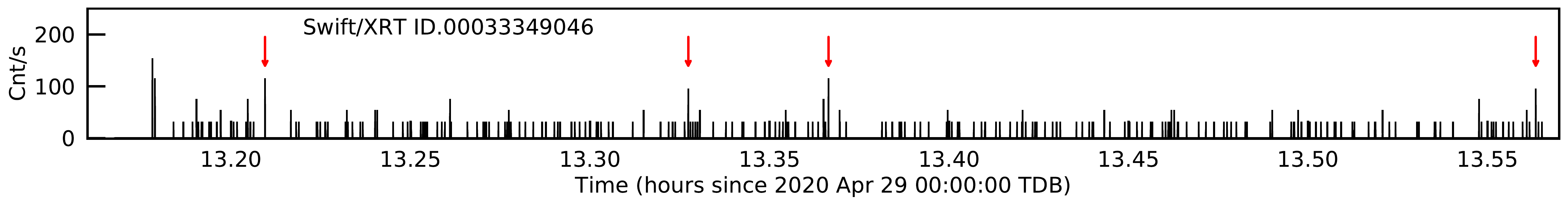}
    \includegraphics[width=1.45\columnwidth]{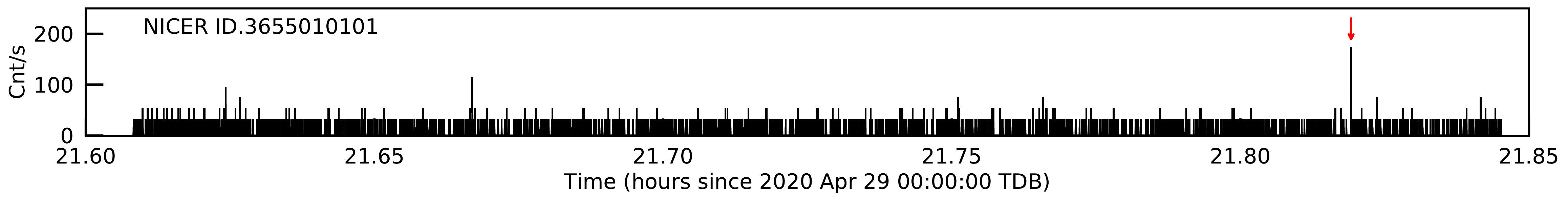}
    \includegraphics[width=1.45\columnwidth]{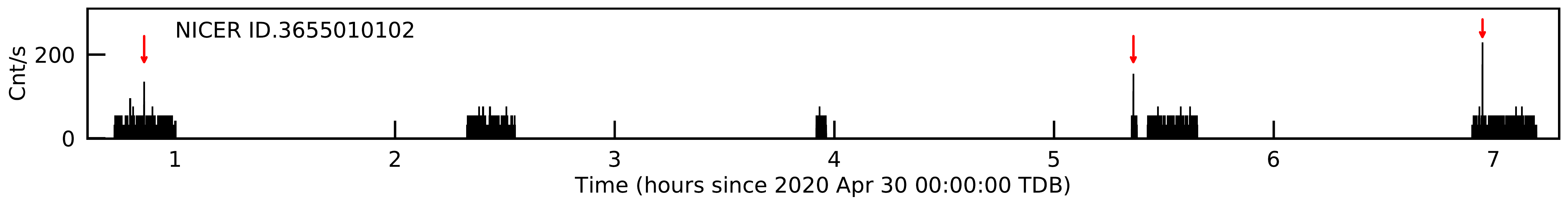}
    \includegraphics[width=1.45\columnwidth]{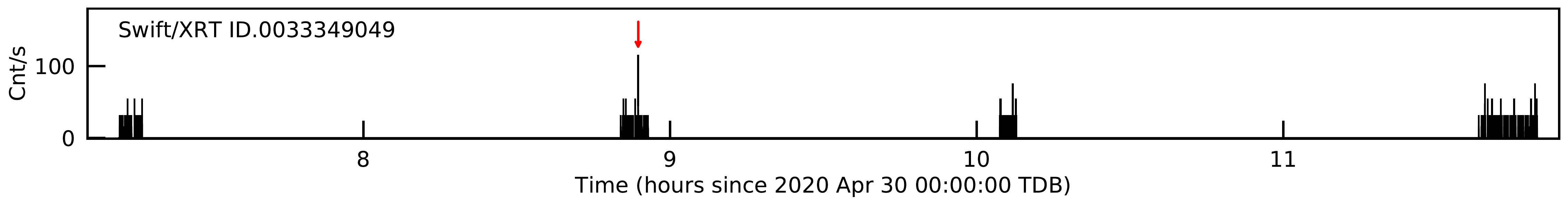}
    \includegraphics[width=1.45\columnwidth]{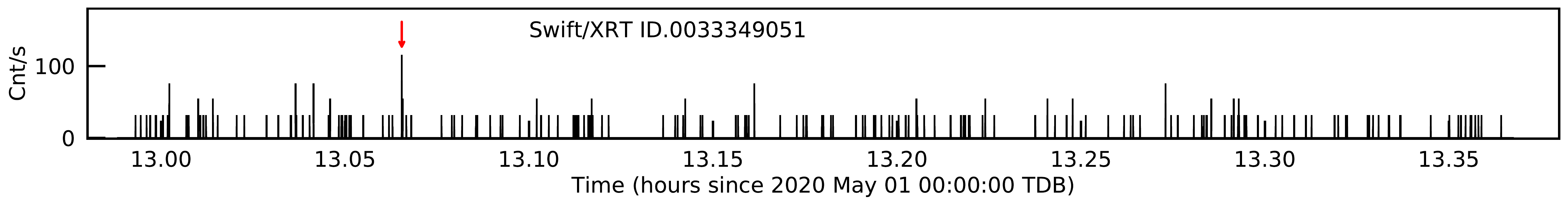}
    \includegraphics[width=1.45\columnwidth]{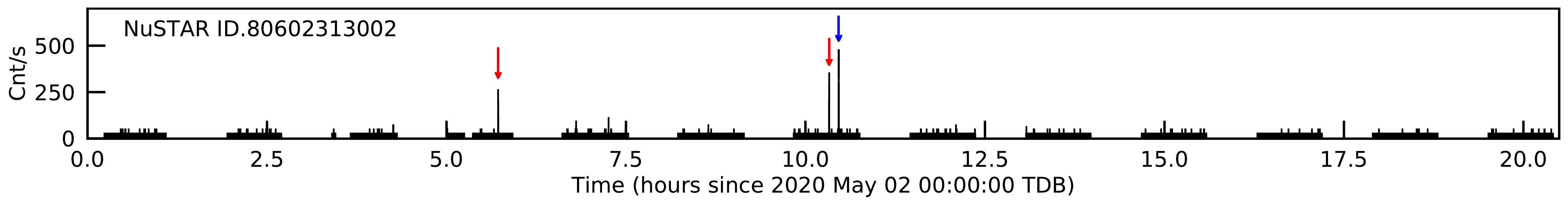}
    \includegraphics[width=1.45\columnwidth]{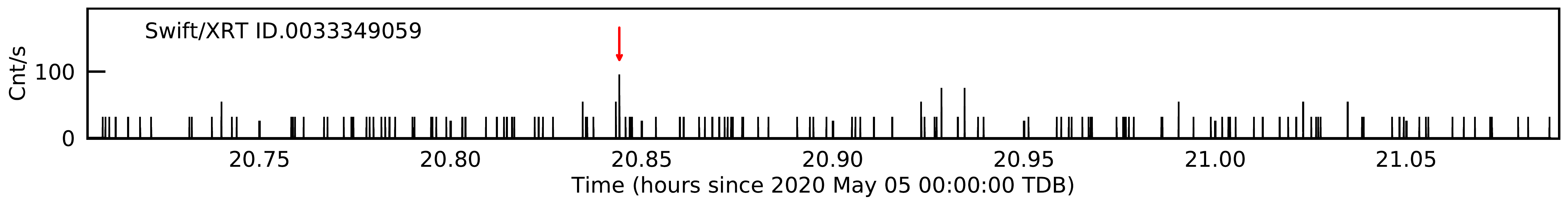}
    \includegraphics[width=1.45\columnwidth]{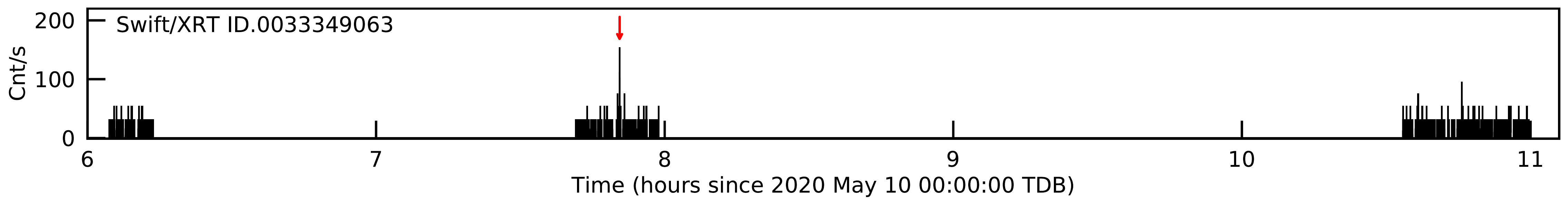}
    \includegraphics[width=1.45\columnwidth]{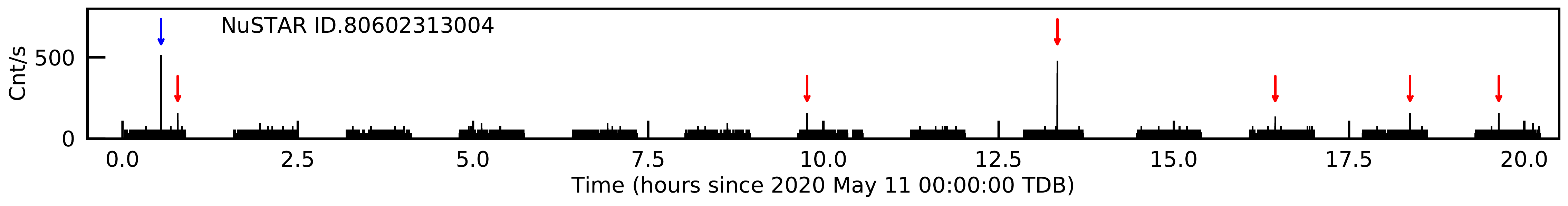}
    \includegraphics[width=1.45\columnwidth]{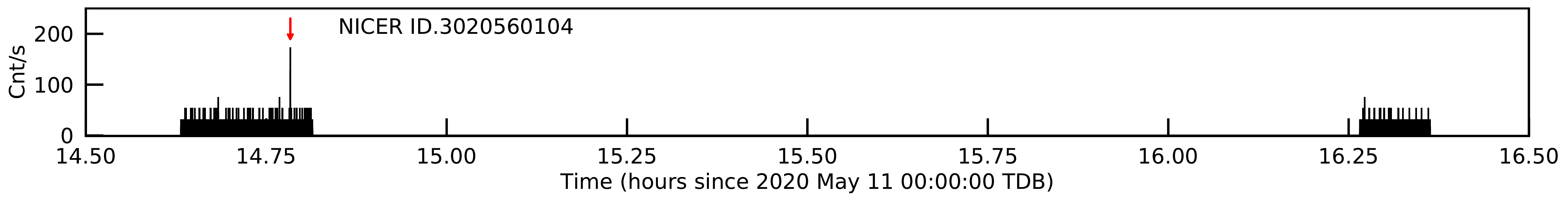}
    \includegraphics[width=1.45\columnwidth]{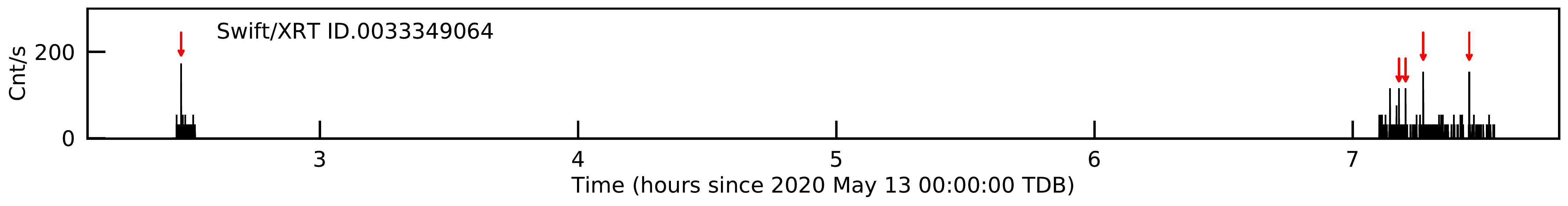}
    \includegraphics[width=1.45\columnwidth]{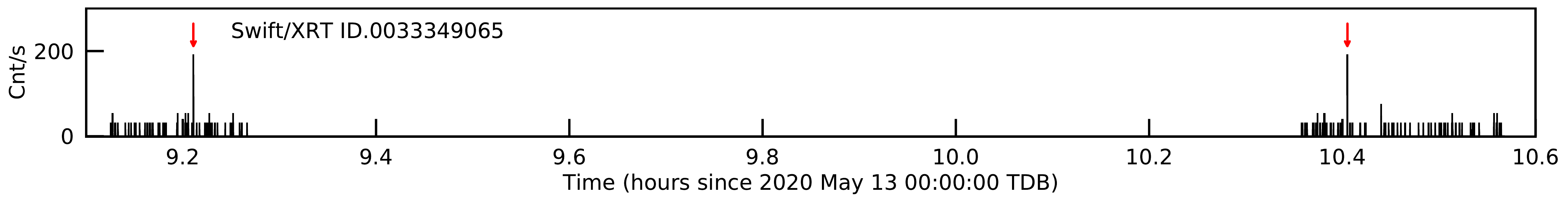}
    \includegraphics[width=1.45\columnwidth]{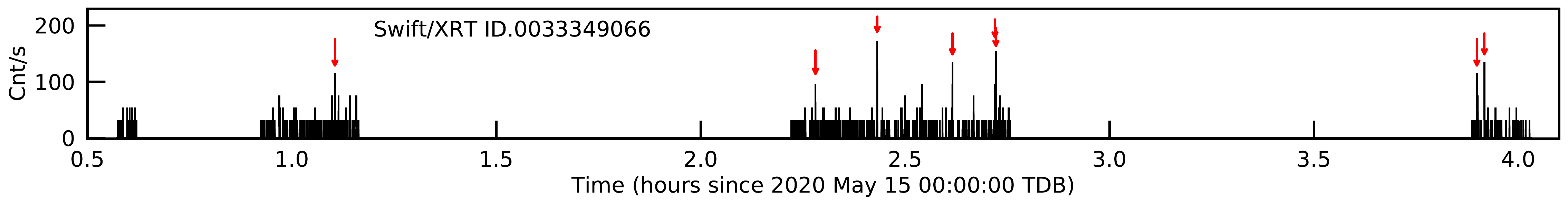}
    \caption{Light curves of \src\ extracted from the \swift/XRT (0.3--10\,keV), \nustar\ (3--79\,keV) and \nicer\ (0.3--10\,keV) data in which we detected bursts. All bursts are marked by arrows (in blue for the two cases for which we performed a spectral analysis). The light curves were binned at 62.5\,ms in all cases except for the data of the first \swift/XRT PC-mode observation (ID. 00968211001), binned at 2.5073\,s.}
    \label{fig:bursts}
\end{figure*}

\subsection{Spectral Analysis}\label{sec:spec}

The spectral analysis was performed with the {\sc xspec} fitting package. We adopted the {\sc Tbabs} model \citep{wilms00} to describe the photoelectric absorption by the interstellar medium. The \nustar\ and \nicer\ background-subtracted spectra were grouped in at least 50 and 20 counts per bin, respectively. The \swift/XRT spectra were grouped according to a minimum number of counts variable from observation to observation. We used the Cash statistic to compute model parameters and their uncertainties.

Fig.\,\ref{fig:spec} shows the spectra extracted from nearly simultaneous \nustar\ and \swift/XRT data. The broad-band spectrum is well described by an absorbed blackbody model plus a power-law component accounting for the emission above 10\,keV. The hydrogen column density was held fixed to \nh\ = $2.3\times10^{22}$\,\cmdue\ in the fits, i.e. the value derived by \cite{cotizelati18} (this is compatible with that given by \citealt{younes17}). For the first epoch (May 2), the best-fitting values are: $kT_{\rm BB} = 0.59^{+0.06}_{-0.05}$\,keV, $R_{\rm BB} = 0.85^{+0.35}_{-0.18}$\,km, and photon index $\Gamma = 1.17 \pm 0.06$ (C-stat = 160.24 for 146 degrees of freedom, dof). For the second epoch (May 11), we derived $kT_{\rm BB} = 0.52 \pm 0.04$\,keV, $R_{\rm BB} =  1.03^{+0.32}_{-0.20}$\,km and $\Gamma = 1.22 \pm 0.06$ (C-stat = 112.17/145 dof). The observed fluxes were (6.9 $\pm$ 0.1) $\times 10^{-12}$ and 5.9$^{+0.3}_{-0.1}$ $\times 10^{-12}$\,\flux\ (0.3--25\,keV), chronologically, giving luminosities of (4.01 $\pm$ 0.08) $\times 10^{34} d^2_{6.6}$ and (3.46 $\pm$ 0.08) $\times 10^{34} d^2_{6.6}$\,\lum, where $d_{6.6}$ is the source distance in units of 6.6\,kpc. At both epochs, the power-law component accounted for $\sim$86\% of the total observed flux and its luminosity varied from (3.23 $\pm$ 0.09) $\times 10^{34} d^2_{6.6}$ to (2.84 $\pm$ 0.07) $\times 10^{34} d^2_{6.6}$\,\lum\ (0.3--25\,keV). 

We then fit the same model to the \swift/XRT spectra jointly, and repeated the same procedure for the \nicer\ spectra (\nh\ was fixed to the above value).  To avoid covariance between the values of the blackbody temperature and normalization due to the limited energy band adopted for \nicer\ spectra (1--5\,keV), we tied up the temperature across all data sets except for the first one. Firstly, we allowed the photon index to vary in the fits. However, we could not obtain meaningful constraints on this parameter over the energy range covered by \swift\ and \nicer. We then repeated the analysis by fixing it to $\Gamma = 1.2$, i.e. the value measured using the \nustar\ observations. We obtained C-stat = 134.95 for 111 dof for the \swift\ data and C-stat = 447.71 for 480 dof for the \nicer\ data. 

The blackbody temperature reached a value of $kT_{\rm BB}= 1.61^{+0.20}_{-0.14}$\,keV  about 75\,min after the first BAT trigger on April 27 at 18:26:20 UT \citep{palmer20}. It decreased to (0.80 $\pm$ 0.02)\,keV in the following day, and attained values in the range 0.45--0.6\,keV over the last $\sim$10 days of our monitoring (Table\,\ref{tab:observations}; Fig.\,\ref{fig:spec}). During the first $\sim$20 days of this new active phase, the observed flux dropped from $5.01^{+0.05}_{-0.59} \times 10^{-11}$ to $2.3^{+0.4}_{-1.0} \times 10^{-12}$\,\flux\ (0.3--10\,keV; Table\,\ref{tab:observations}; Fig.\,\ref{fig:spec}). These values translate into a luminosity of $(3.2 \pm 0.3) \times 10^{35} d^2_{6.6}$ and $(2.5 \pm 0.2) \times 10^{34} d^2_{6.6}$\,\lum\ (0.3--10\,keV), respectively. An XRT observation performed on April 23 (only 4 days prior to the outburst onset) found \src\ in quiescence with net count rate of $0.012 \pm 0.002$ counts s$^{-1}$ (0.3--10\,keV), corresponding to an observed flux of $\sim4.5 \times 10^{-13}$\,\flux\ and a luminosity of $\sim 5.8 \times 10^{33} d^2_{6.6}$\,\lum\ (assuming an absorbed blackbody spectrum with $kT_{\rm BB} = 0.5$\,keV, \nh\ $= 2.3 \times 10^{22}$\,\cmdue). 

\subsection{Burst Search and Properties}

We inspected the light curves of all observations for the presence of short bursts.  
Our search algorithm estimates the Poisson probability for an event to be a random fluctuation compared to the average number of counts per bin in the full observation, considering the total number of time bins $N$. We applied this algorithm to light curves binned with different time resolutions (2$^{-4}$, 2$^{-5}$ and 2$^{-6}$\,s) to be sensitive to bursts of different duration, except for the \swift/XRT PC-mode event files that were binned at the available timing resolution (2.5073\,s). Bins having a probability smaller than 10$^{-4}(N N_{\rm trials})^{-1}$ are identified as bursts ($N_{\rm trials}$ is the number of different time resolutions adopted for the search). 
In Table\,\ref{tab:bursts}, we report the epochs of the bursts referred to the solar system barycenter. Fluence and duration are given for the bursts detected in the \nicer\ and \nustar\ data sets. Their light curves are shown in Fig.\,\ref{fig:bursts}. 
We do not report on the $\sim$25 short bursts detected in the first \nicer\ observation (Obs. ID 3020560101; see Table 1 of \citealt{younes20}) due to the complex light curve and instrument saturation problems.

We extracted the spectra only for those events with at least 30 net counts, that is, two bursts detected in the \nustar\ observations (80602313002 \#3 and 80602313004 \#1 in Table\,\ref{tab:bursts}). We fitted the spectra using single-component models (a power-law, a blackbody and an optically thin
thermal bremsstrahlung). The blackbody and power-law model fits gave a satisfactory description for both events with a goodness probability\footnote{\url{https://heasarc.gsfc.nasa.gov/xanadu/xspec/manual/node84.html}} of $\sim$55\% and $\sim$40\%, respectively.  
For the blackbody model, we derived a temperature equivalent to (2.9 $\pm$ 0.5)\,keV for 80602313002 \#3 and (3.9 $\pm$ 0.7)\,keV for 80602313004 \#1. The corresponding fluxes were (1.0 $\pm$ 0.3) $\times$ 10$^{-8}$ and (1.8 $\pm$ 0.6) $\times$ 10$^{-8}$\,\flux\ in the 3--79\,keV energy range, converting to luminosity of (5.3 $\pm$ 1.5) $\times$ 10$^{37} d^2_{6.6}$ and (9.3 $\pm$ 3.1) $\times$ 10$^{37} d^2_{6.6}$\,\lum.

\section{Discussion} \label{sec:discuss}
 
Since its discovery in 2014, the magnetar \src\ has been a prolific source, showing numerous X-ray outbursts and frequent bursting activity. We presented here the results of an intensive X-ray monitoring campaign of this source over about three weeks since the end of April 2020, when it emitted a forest of X-ray bursts, and two bright radio millisecond bursts with characteristics strongly reminiscent of FRBs \citep{scholz20,bochenek20}. \\ 

\noindent
1. {\it Spin period and pulse profiles}. We detected the source spin period in the combined \nicer\ data sets acquired on April 29--30, and in both \nustar\ data sets on May 2 and 11.  Unfortunately, the spacing between the few detections, and the uncertainties on the periods, prevented us from extracting a phase-connected timing solution. The spin period measurements at the different epochs allowed us to set an upper limit on the period derivative of $|\dot{P}| < 3 \times 10^{-11}$\,s\,s$^{-1}$ (at 3$\sigma$ c.l.). This limit is compatible with the spin-down rate of $\dot{P} \sim 1.43 \times 10^{-11}$\,\ss\ derived by \cite{israel16} in 2014, using a phase-connected timing analysis. 

The double-peaked morphology of the \nustar\ pulse profiles is markedly different from the quasi-sinusoidal modulation observed in the \nicer\ observation a few days before and in previous X-ray observations of the source \citep{israel16}. 
Timing noise and large pulse profile changes (in time and energy) are common during magnetar outbursts (e.g. \citealt{dib14}; \citealt{esposito18} for a review, and references therein), especially following X-ray bursting activity. The magnetar magnetosphere is subject to rapid changes before setting to a new quiescent configuration, which are responsible for the fast profile variations especially in the hard X-rays, where the emission is dominated by nonthermal photons. These changes might also lead to the formation of new bundles and hot spots on the surface, modifying the pulse profile also in the soft X-ray range. \\

\noindent
2. {\it Luminosity, spectral evolution, and bursting activity}. About three days before its reactivation, \src\, was observed by \swift/XRT at a luminosity of $\sim5.8\times 10^{33} d^2_{6.6}$\,\lum\ (0.3--10\,keV). Following the source reactivation, the X-ray luminosity reached a peak value of $\sim$3.2 $\times$ 10$^{35} d^2_{6.6}$\,\lum, making this event the most powerful outburst detected from \src\ so far. The luminosity then dropped by more than one order of magnitude, down to $\sim$2.5 $\times$ 10$^{34} d^2_{6.6}$\,\lum\ about 3 weeks later. However, this is still a factor $\sim$4 larger than the pre-outburst level. A similar rapid decay pattern was also observed for the strong outbursts in May and June 2016 \citep{younes17} and, overall, is not uncommon for magnetars in outburst \citep{cotizelati18}.

\begin{figure*}
  \centering
    \includegraphics[width=14cm]{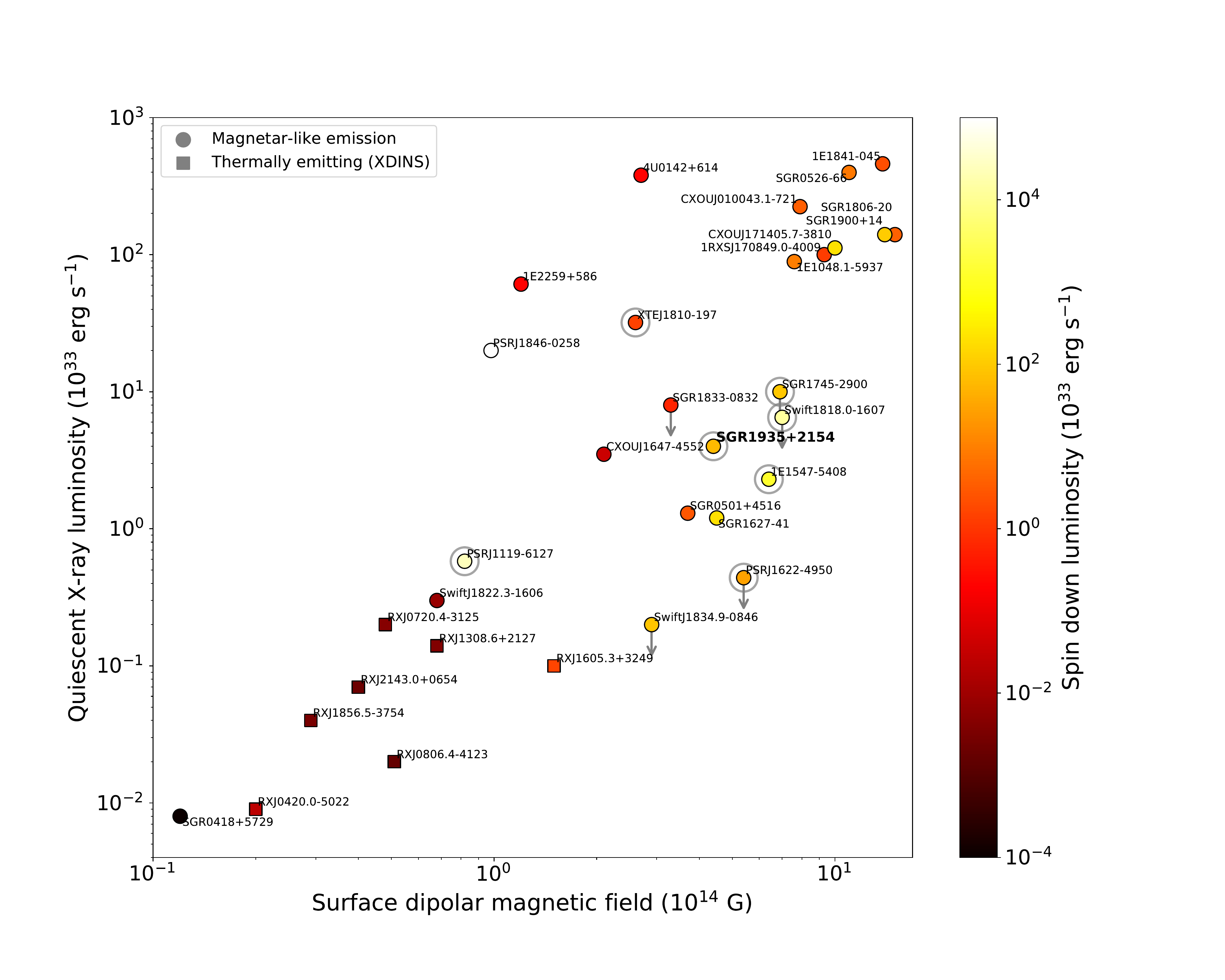}
    \caption{Quiescent X-ray luminosity of magnetars as a function of their dipolar magnetic field at the pole. Circles denote radio-loud magnetars, either in the form of bursts (\src; in bold) or pulsed emission (other sources). Markers are color-coded according to the spin-down power of each source. Values are from the Magnetar Outburst Online Catalogue (\url{http://magnetars.ice.csic.es/}; \citealt{cotizelati18}), with updates for \psr, \sgra\ and \coronamag\ \citep{camilo18,ybk17,esposito20}.}
    \label{fig:discussion}
\end{figure*}

During the entire monitoring, \src\ showed a thermal spectrum in the soft X-rays well described by an absorbed blackbody model quickly cooling from a temperature of $\sim$1.6\,keV to $\sim$0.45--0.6\,keV. 
Emission was detected up to $\sim$25\,keV in the \nustar\ observations. The spectral shape was identical at the two epochs, and was adequately modelled by a power-law model with index $\Gamma \sim 1.2$ and luminosity $\sim4 \times 10^{34} d^2_{6.6}$\,\lum\ (extrapolated to the 10--50\,keV energy range). Hard X-ray emission from \src\ was seen also in a \nustar\ pointing performed $\sim$ 5 days after the 2015 outburst onset. In that case, the high-energy spectrum could be described by a slightly harder power-law component ($\Gamma$ $\sim$ 0.9) with a lower luminosity, $\sim$1 $\times$ 10$^{34} d^2_{6.6}$\,\lum\ (10--50\,keV; \citealt{younes17}).
However, the spectral evolution during this last outburst is different from that observed in the previous events, where 
the luminosity decay could be ascribed to the evolution of the high-energy component \citep{younes17}. 

The bursting activity of \src\ during this new outburst is not dissimilar from that previously observed in this and other magnetars. However, such activity is not so prolific in all magnetars, and it is expected to depend on the age of the source and the tangled configuration of its magnetic field \citep{perna11,vigano13}. A very rough proxy for it is provided by the quiescent X-ray luminosity, which is predicted to be higher in magnetars with a more tangled and powerful magnetic field in the crust, since they are subject to larger crustal currents and $B$-field crustal dissipation (see Fig.\,\ref{fig:discussion}). A significant anti--correlation between magnetar quiescent luminosities and their luminosity increases in outburst was observed \citep{pons12,cotizelati18}, suggesting the existence of a limiting luminosity of $\sim 10^{36}$\lum\ for magnetar outbursts (regardless of the source quiescent level), which holds also for the case of \src .\\

\noindent
3. {\it Comparison with other magnetars and FRBs}. Comparing the short X-ray bursts and outburst emitted by \src\ with those of the other Galactic magnetars, they are perfectly in line with expectations. There is nothing in the X-ray emission properties of this magnetar that would make it peculiar in any aspect \citep{cotizelati18}. However, the simultaneous detection of radio bursts with a bright magnetar-like burst \citep{scholz20,bochenek20,mereghetti20} showed for the first time that magnetar bursts might have bright radio counterparts. This result is particularly interesting in the context of the physical interpretation of FRBs, bright ms-duration transients from distant galaxies. 
Their brightness temperatures imply a coherent radio emission, suggesting a connection with pulsar emission mechanisms.
Several repeating FRBs have been discovered \citep{spitler14,spitler16}, reinforcing their proposed interpretation in terms of young bursting magnetars in other galaxies (e.g., \citealt{popov13, margalit20}, and references therein).

Radio pulsed emission was so far restricted to five magnetars (see Fig.\,\ref{fig:discussion}). Such emission is at variance with the typical radio pulsar emission, and it is always connected to some extent with the magnetar X-ray activation. However, similarly to radio pulsars, all radio-loud magnetars have a large spin-down power compared to their radio-quiet siblings, and quiescent X-ray luminosity below their rotational power (with the exception of \xte; \citealt{rea12b,cotizelati18} and Fig.\,\ref{fig:discussion}). \src\ has a high rotational power, but so far it did not show any radio pulsations \citep{younes17,lin20b}, while surprisingly emitting radio bursts during the outburst we report here. From the study of the bursting activity of this source, it becomes clear that: 1) not all X-ray magnetar bursts have necessarily a radio counterpart (see also \citealt{archibald20}), and 2) many radio bursts from magnetars might have been missed due to the lack of large field-of-view instruments in the radio band. Hence, it might be a common characteristic after all. Future detections will shed light on these ms-radio bursts, their connection (or not) with faint radio pulsations (i.e. bright single pulses), their preferred X-ray burst counterparts. Population synthesis studies will allow a comparison between their rates and luminosity distributions and those observed in FRBs.

\begin{deluxetable*}{cccccc}[htb]
\tabletypesize{\footnotesize}
\tablecaption{Log of X-ray bursts detected in all datasets except for the first \nicer\ observation (ID: 3020560101; see text for details). 
\label{tab:bursts}}
\tablecolumns{4}
\tablenum{2}
\tablewidth{0pt}
\tablehead{
\colhead{Instrument} &
\colhead{Obs.ID\tablenotemark{\footnotesize a}} &
\colhead{Burst epoch} &
\colhead{Fluence\tablenotemark{\footnotesize b}} &
\colhead{Duration\tablenotemark{\footnotesize c}}\\
\colhead{} &
\colhead{} & 
\colhead{YYYY-MM-DD hh:mm:ss (TDB)} &
\colhead{(counts)} & 
\colhead{(ms)} 
}
\startdata
\swift/XRT (PC)\tablenotemark{\footnotesize d} & 00968211001 \#1 & 2020-04-27 19:46:52 & --  & --  \\
                 &   \#2           & 19:57:22            & -- & --  \\    
                 &   \#3          & 20:03:38            & --  & --   \\    
                 &   \#4          & 20:09:12            & --  & --   \\                  
                 &   \#5          & 20:14:42            & --  & --   \\   
                 &   \#6          & 20:15:30            & --  & --   \\  
\swift/XRT (WT)\tablenotemark{\footnotesize d} & 00033349046 \#1 & 13:12:34  & --  & --   \\
                                               &      \#2        & 13:19:38  & --  & --   \\
                                               &      \#3        & 13:21:59  & --  & --   \\
                                               &      \#4        & 13:33:48  & --  & --   \\
\nicer/XTI       & 3655010101 \#1 & 2020-04-29 21:49:09 & 8 & 62.5  \\
\nicer/XTI       & 3655010102 \#1 & 2020-04-30 00:51:27  & 6 & 31.25  \\
                 &    \#2         &  05:21:38.89        & 7 & 62.5 \\
                 &    \#3         &    06:56:59         & 11 & 62.5  \\
\swift/XRT (WT)\tablenotemark{\footnotesize d} & 00033349049 \#1 & 2020-04-30 08:53:44 & --  & --   \\
\swift/XRT (WT)\tablenotemark{\footnotesize d} & 00033349051 \#1 & 2020-05-01 13:03:55 & --  & --   \\
\nustar\         & 80602313002 \#1 & 2020-05-02 05:43:12 & 11 & 31.25 \\
                 &      \#2      &  10:19:46           & 13 & 31.25  \\
                 &      \#3       &  10:27:46           & 46 & 0.125 \\
\swift/XRT (WT)\tablenotemark{\footnotesize d} & 00033349059 \#1 & 2020-05-05 20:50:38  & --  & --   \\
\swift/XRT (WT)\tablenotemark{\footnotesize d} & 00033349063 \#1 & 2020-05-10 07:50:37  & --  & --   \\
\nustar\         & 80602313004 \#1 & 2020-05-11 00:33:00 & 43 & 93.75  \\
                 &      \#2       & 00:47:14 & 7 & 62.5  \\
                 &      \#3       & 09:46:00 & 7 & 62.5  \\
                 &      \#4       & 13:20:16 & 25 & 46.875  \\
                 &      \#5       & 18:22:15 & 7 & 62.5  \\
                 &      \#6       & 19:38:10 & 7 & 62.5  \\
\nicer/XTI       & 3020560104 \#1 & 2020-05-11 14:47:00 & 8 & 62.5 \\                 
\swift/XRT (WT)\tablenotemark{\footnotesize d} & 00033349064 \#1 & 2020-05-13 02:27:45 & --  & --   \\
                 &      \#2       & 07:10:44           & --  & --   \\
                 &      \#3      & 07:12:16         & --  & --   \\
                 &      \#4       & 07:16:22        & --  & --   \\
                 &      \#5      & 07:27:04         & --  & --   \\
\swift/XRT (WT)\tablenotemark{\footnotesize d} & 00033349065 \#1 & 2020-05-13 09:12:39  & --  &   \\
                 &      \#2       & 10:24:18        & --  &   \\ 
\swift/XRT (WT)\tablenotemark{\footnotesize d} & 00033349066 \#1 & 2020-05-15 01:06:18  & --  &   \\
                 &      \#2       & 02:16:50        & --  & --   \\
                 &      \#3       & 02:25:54        & --  & --   \\
                 &      \#4       & 02:36:56        & --  & --   \\
                 &      \#5       & 02:43:10        & --  & --   \\
                 &      \#6       & 02:43:19        & --  & --   \\ 
                 &      \#7       & 03:53:54        & --  & --   \\
                 &      \#8       & 03:55:00        & --  & --   \\
\enddata

\tablenotetext{a}{The notation \#N corresponds to the burst number in a given observation.}
\tablenotetext{b}{The fluence refers to the 0.3--10\,keV range for \nicer/XTI and 3--79\,keV for \nustar.}
\tablenotetext{c}{The duration has to be considered as an approximate value. We estimated it by summing the 15.625-ms time bins showing enhanced emission for the structured bursts, and by setting it equal to the coarser time resolution at which the burst is detected in all the other cases.}
\tablenotetext{d}{Fluence and maximum durations are not reported for the bursts detected by \swift/XRT owing to uncertainties related to the detector saturation limits.}
\end{deluxetable*}

\acknowledgments
This research is based on observations with \nicer\ (NASA), \nustar\ (CaltTech/NASA/JPL) and \swift\ (NASA/ASI/UK). We thank the \nicer, \nustar\ and \swift\ teams for promptly scheduling our observations. We made use of public \swift\, and \nicer\, observations asked by the CHIME/FRB collaboration and the \nicer\, team, that we acknowledge.
We thank Roberto Turolla and Andrea Possenti for useful suggestions.
We acknowledge support from the PHAROS COST Action (CA16214). A.B. and F.C.Z. are supported by Juan de la Cierva fellowships. A.B., F.C.Z., and N.R. are supported by grants SGR2017-1383, PGC2018-095512-BI00, and the ERC Consolidator grant ``MAGNESIA" (No. 817661). G.L.I., S.M. and A.T. acknowledge financial support from the Italian MIUR through PRIN grant 2017LJ39LM. G.L.I. also acknowledges funding from ASI-INAF agreements I/037/12/0 and 2017-14-H.O.
\facilities{\nicer, \nustar, \swift\ (XRT).}
\software{HEASoft (v6.27.2), FTOOLS (v6.27; \citealt{blackburn95}), XSPEC (v12.11.0h; \citealt{arnaud96}), NICERDAS (v7a), NuSTARDAS (v1.9.2), MATPLOTLIB (v3.2.1; \citealt{hunter07}), NUMPY (v1.18.4; \citealt{vanderWalt11}).}
 
\bibliographystyle{aasjournal}
\bibliography{biblio}

\begin{thebibliography}{}
\expandafter\ifx\csname natexlab\endcsname\relax\def\natexlab#1{#1}\fi

\bibitem[{{Archibald} {et~al.}(2016){Archibald}, {Kaspi}, {Tendulkar}, \&
  {Scholz}}]{archibald16}
{Archibald}, R.~F., {Kaspi}, V.~M., {Tendulkar}, S.~P., \& {Scholz}, P. 2016,
  \apjl, 829, L21

\bibitem[{{Archibald} {et~al.}(2020){Archibald}, {Scholz}, {Kaspi},
  {Tendulkar}, \& {Beardmore}}]{archibald20}
{Archibald}, R.~F., {Scholz}, P., {Kaspi}, V.~M., {Tendulkar}, S.~P., \&
  {Beardmore}, A.~P. 2020, \apj, 889, 160

\bibitem[{{Arnaud}(1996)}]{arnaud96}
{Arnaud}, K.~A. 1996, in Astronomical Data Analysis Software and Systems V,
  Vol. 101, XSPEC: The First Ten Years, ed. G.~H. {Jacoby} \& J.~{Barnes} (ASP,
  San Francisco), 17--20

\bibitem[{{Blackburn}(1995)}]{blackburn95}
{Blackburn}, J.~K. 1995, in Astronomical Data Analysis Software and Systems
  IV., Vol.~77, {FTOOLS: A FITS Data Processing and Analysis Software Package},
  ed. R.~A. {Shaw}, H.~E. {Payne}, \& J.~J.~E. {Hayes} (ASP Conf. Ser., San
  Francisco, CA), 367

\bibitem[{{Bochenek} {et~al.}(2020){Bochenek}, {Ravi}, {Belov}, {Hallinan},
  {Kocz}, {Kulkarni}, \& {McKenna}}]{bochenek20}
{Bochenek}, C.~D., {Ravi}, V., {Belov}, K.~V., {et~al.} 2020, arXiv e-prints,
  arXiv:2005.10828

\bibitem[{{Borghese} {et~al.}(2018){Borghese}, {Coti Zelati}, {Esposito},
  {Rea}, {De Luca}, {Bachetti}, {Israel}, {Perna}, \& {Pons}}]{borghese18}
{Borghese}, A., {Coti Zelati}, F., {Esposito}, P., {et~al.} 2018, \mnras, 478,
  741

\bibitem[{{Burrows} {et~al.}(2005){Burrows}, {Hill}, {Nousek}, {Kennea},
  {Wells}, {Osborne}, {Abbey}, {Beardmore}, {Mukerjee}, {Short}, {Chincarini},
  {Campana}, {Citterio}, {Moretti}, {Pagani}, {Tagliaferri}, {Giommi},
  {Capalbi}, {Tamburelli}, {Angelini}, {Cusumano}, {Br{\"a}uninger}, {Burkert},
  \& {Hartner}}]{burrows05}
{Burrows}, D.~N., {Hill}, J.~E., {Nousek}, J.~A., {et~al.} 2005, Space Science
  Reviews, 120, 165

\bibitem[{{Camilo} {et~al.}(2018){Camilo}, {Scholz}, {Serylak}, {Buchner},
  {Merryfield}, {Kaspi}, {Archibald}, {Bailes}, {Jameson}, {van Straten},
  {Sarkissian}, {Reynolds}, {Johnston}, {Hobbs}, {Abbott}, {Adam}, {Adams},
  {Alberts}, {Andreas}, {Asad}, {Baker}, {Baloyi}, {Bauermeister}, {Baxana},
  {Bennett}, {Bernardi}, {Booisen}, {Booth}, {Botha}, {Boyana}, {Brederode},
  {Burger}, {Cheetham}, {Conradie}, {Conradie}, {Davidson}, {De Bruin}, {de
  Swardt}, {de Villiers}, {de Villiers}, {de Villiers}, {de Villiers}, {De
  Waal}, {Dikgale}, {du Toit}, {du Toit}, {Esterhuyse}, {Fanaroff}, {Fataar},
  {Foley}, {Foster}, {Fourie}, {Gamatham}, {Gatsi}, {Geschke}, {Goedhart},
  {Grobler}, {Gumede}, {Hlakola}, {Hokwana}, {Hoorn}, {Horn}, {Horrell},
  {Hugo}, {Isaacson}, {Jacobs}, {Jansen van Rensburg}, {Jonas}, {Jordaan},
  {Joubert}, {Joubert}, {J{\'o}zsa}, {Julie}, {Julius}, {Kapp}, {Karastergiou},
  {Karels}, {Kariseb}, {Karuppusamy}, {Kasper}, {Knox-Davies}, {Koch},
  {Kotz{\'e}}, {Krebs}, {Kriek}, {Kriel}, {Kusel}, {Lamoor}, {Lehmensiek},
  {Liebenberg}, {Liebenberg}, {Lord}, {Lunsky}, {Mabombo}, {Macdonald},
  {Macfarlane}, {Madisa}, {Mafhungo}, {Magnus}, {Magozore}, {Mahgoub}, {Main},
  {Makhathini}, {Malan}, {Malgas}, {Manley}, {Manzini}, {Marais}, {Marais},
  {Marais}, {Maree}, {Martens}, {Matshawule}, {Matthysen}, {Mauch}, {McNally},
  {Merry}, {Millenaar}, {Mjikelo}, {Mkhabela}, {Mnyand u}, {Moeng}, {Mokone},
  {Monama}, {Montshiwa}, {Moss}, {Mphego}, {New}, {Ngcebetsha}, {Ngoasheng},
  {Niehaus}, {Ntuli}, {Nzama}, {Obies}, {Obrocka}, {Ockards}, {Olyn}, {Oozeer},
  {Otto}, {Padayachee}, {Passmoor}, {Patel}, {Paula}, {Peens-Hough},
  {Pholoholo}, {Prozesky}, {Rakoma}, {Ramaila}, {Rammala}, {Ramudzuli},
  {Rasivhaga}, {Ratcliffe}, {Reader}, {Renil}, {Richter}, {Robyntjies},
  {Rosekrans}, {Rust}, {Salie}, {Sambu}, {Schollar}, {Schwardt}, {Seranyane},
  {Sethosa}, {Sharpe}, {Siebrits}, {Sirothia}, {Slabber}, {Smirnov}, {Smith},
  {Sofeya}, {Songqumase}, {Spann}, {Stappers}, {Steyn}, {Steyn}, {Strong},
  {Struthers}, {Stuart}, {Sunnylall}, {Swart}, {Taljaard}, {Tasse}, {Taylor},
  {Theron}, {Thondikulam}, {Thorat}, {Tiplady}, {Toruvanda}, {van Aardt}, {van
  Balla}, {van den Heever}, {van der Byl}, {van der Merwe}, {van der Merwe},
  {van Niekerk}, {van Rooyen}, {van Staden}, {van Tonder}, {van Wyk}, {Wait},
  {Walker}, {Wallace}, {Welz}, {Williams}, {Xaia}, {Young}, \&
  {Zitha}}]{camilo18}
{Camilo}, F., {Scholz}, P., {Serylak}, M., {et~al.} 2018, \apj, 856, 180

\bibitem[{{Collazzi} {et~al.}(2015){Collazzi}, {Kouveliotou}, {van der Horst},
  {Younes}, {Kaneko}, {G{\"o}{\u{g}}{\"u}{\textcommabelow s}}, {Lin}, {Granot},
  {Finger}, {Chaplin}, {Huppenkothen}, {Watts}, {von Kienlin}, {Baring},
  {Gruber}, {Bhat}, {Gibby}, {Gehrels}, {McEnery}, {van der Klis}, \&
  {Wijers}}]{collazzi15}
{Collazzi}, A.~C., {Kouveliotou}, C., {van der Horst}, A.~J., {et~al.} 2015,
  \apjs, 218, 11

\bibitem[{{Cordes} \& {Chatterjee}(2019)}]{cordes19}
{Cordes}, J.~M., \& {Chatterjee}, S. 2019, \araa, 57, 417

\bibitem[{{Coti Zelati} {et~al.}(2018){Coti Zelati}, {Rea}, {Pons}, {Campana},
  \& {Esposito}}]{cotizelati18}
{Coti Zelati}, F., {Rea}, N., {Pons}, J.~A., {Campana}, S., \& {Esposito}, P.
  2018, \mnras, 474, 961

\bibitem[{{D'A{\`i}} {et~al.}(2016){D'A{\`i}}, {Evans}, {Burrows}, {Kuin},
  {Kann}, {Campana}, {Maselli}, {Romano}, {Cusumano}, {La Parola}, {Barthelmy},
  {Beardmore}, {Cenko}, {De Pasquale}, {Gehrels}, {Greiner}, {Kennea}, {Klose},
  {Melandri}, {Nousek}, {Osborne}, {Palmer}, {Sbarufatti}, {Schady}, {Siegel},
  {Tagliaferri}, {Yates}, \& {Zane}}]{dai16}
{D'A{\`i}}, A., {Evans}, P.~A., {Burrows}, D.~N., {et~al.} 2016, \mnras, 463,
  2394

\bibitem[{{Dib} \& {Kaspi}(2014)}]{dib14}
{Dib}, R., \& {Kaspi}, V.~M. 2014, \apj, 784, 37

\bibitem[{{Esposito} {et~al.}(2018){Esposito}, {Rea}, \& {Israel}}]{esposito18}
{Esposito}, P., {Rea}, N., \& {Israel}, G.~L. 2018, in Timing Neutron Stars:
  Pulsations, Oscillations and Explosions, ed. T.~{Belloni}, M.~{Mendez}, \&
  C.~{Zhang}, ASSL, Springer, in press (preprint: astro-ph/1803.05716)

\bibitem[{{Esposito} {et~al.}(2020){Esposito}, {Rea}, {Borghese}, {Coti
  Zelati}, {Vigan{\`o}}, {Israel}, {Tiengo}, {Ridolfi}, {Possenti}, {Burgay},
  {G{\"o}tz}, {Pintore}, {Stella}, {Dehman}, {Ronchi}, {Campana},
  {Garcia-Garcia}, {Graber}, {Mereghetti}, {Perna}, {Rodr{\'\i}guez Castillo},
  {Turolla}, \& {Zane}}]{esposito20}
{Esposito}, P., {Rea}, N., {Borghese}, A., {et~al.} 2020, \apjl, 896, L30

\bibitem[{{Gavriil} {et~al.}(2008){Gavriil}, {Gonzalez}, {Gotthelf}, {Kaspi},
  {Livingstone}, \& {Woods}}]{gavriil08}
{Gavriil}, F.~P., {Gonzalez}, M.~E., {Gotthelf}, E.~V., {et~al.} 2008, Science,
  319, 1802

\bibitem[{{Gehrels} {et~al.}(2004){Gehrels}, {Chincarini}, {Giommi}, {Mason},
  {Nousek}, {Wells}, {White}, {Barthelmy}, {Burrows}, {Cominsky}, {Hurley},
  {Marshall}, {M{\'e}sz{\'a}ros}, {Roming}, {Angelini}, {Barbier}, {Belloni},
  {Campana}, {Caraveo}, {Chester}, {Citterio}, {Cline}, {Cropper}, {Cummings},
  {Dean}, {Feigelson}, {Fenimore}, {Frail}, {Fruchter}, {Garmire}, {Gendreau},
  {Ghisellini}, {Greiner}, {Hill}, {Hunsberger}, {Krimm}, {Kulkarni}, {Kumar},
  {Lebrun}, {Lloyd-Ronning}, {Markwardt}, {Mattson}, {Mushotzky}, {Norris},
  {Osborne}, {Paczynski}, {Palmer}, {Park}, {Parsons}, {Paul}, {Rees},
  {Reynolds}, {Rhoads}, {Sasseen}, {Schaefer}, {Short}, {Smale}, {Smith},
  {Stella}, {Tagliaferri}, {Takahashi}, {Tashiro}, {Townsley}, {Tueller},
  {Turner}, {Vietri}, {Voges}, {Ward}, {Willingale}, {Zerbi}, \&
  {Zhang}}]{gehrels04}
{Gehrels}, N., {Chincarini}, G., {Giommi}, P., {et~al.} 2004, \apj, 611, 1005

\bibitem[{{Gendreau} {et~al.}(2012){Gendreau}, {Arzoumanian}, \&
  {Okajima}}]{gendreau12}
{Gendreau}, K.~C., {Arzoumanian}, Z., \& {Okajima}, T. 2012, in Society of
  Photo-Optical Instrumentation Engineers (SPIE) Conference Series, Vol. 8443,
  \procspie, 844313

\bibitem[{{Harrison} {et~al.}(2013){Harrison}, {Craig}, {Christensen},
  {Hailey}, {Zhang}, {Boggs}, {Stern}, {Cook}, {Forster}, {Giommi},
  {Grefenstette}, {Kim}, {Kitaguchi}, {Koglin}, {Madsen}, {Mao}, {Miyasaka},
  {Mori}, {Perri}, {Pivovaroff}, {Puccetti}, {Rana}, {Westergaard}, {Willis},
  {Zoglauer}, {An}, {Bachetti}, {Barri{\`e}re}, {Bellm}, {Bhalerao},
  {Brejnholt}, {Fuerst}, {Liebe}, {Markwardt}, {Nynka}, {Vogel}, {Walton},
  {Wik}, {Alexander}, {Cominsky}, {Hornschemeier}, {Hornstrup}, {Kaspi},
  {Madejski}, {Matt}, {Molendi}, {Smith}, {Tomsick}, {Ajello}, {Ballantyne},
  {Balokovi{\'c}}, {Barret}, {Bauer}, {Blandford}, {Brandt}, {Brenneman},
  {Chiang}, {Chakrabarty}, {Chenevez}, {Comastri}, {Dufour}, {Elvis}, {Fabian},
  {Farrah}, {Fryer}, {Gotthelf}, {Grindlay}, {Helfand}, {Krivonos}, {Meier},
  {Miller}, {Natalucci}, {Ogle}, {Ofek}, {Ptak}, {Reynolds}, {Rigby},
  {Tagliaferri}, {Thorsett}, {Treister}, \& {Urry}}]{harrison13}
{Harrison}, F.~A., {Craig}, W.~W., {Christensen}, F.~E., {et~al.} 2013, \apj,
  770, 103

\bibitem[{Hunter(2007)}]{hunter07}
Hunter, J.~D. 2007, Computing in Science \& Engineering, 9, 90

\bibitem[{{Israel} \& {Stella}(1996)}]{israel96}
{Israel}, G.~L., \& {Stella}, L. 1996, \apj, 468, 369

\bibitem[{{Israel} {et~al.}(2008){Israel}, {Romano}, {Mangano}, {Dall'Osso},
  {Chincarini}, {Stella}, {Campana}, {Belloni}, {Tagliaferri}, {Blustin},
  {Sakamoto}, {Hurley}, {Zane}, {Moretti}, {Palmer}, {Guidorzi}, {Burrows},
  {Gehrels}, \& {Krimm}}]{israel08}
{Israel}, G.~L., {Romano}, P., {Mangano}, V., {et~al.} 2008, \apj, 685, 1114

\bibitem[{{Israel} {et~al.}(2016){Israel}, {Esposito}, {Rea}, {Coti Zelati},
  {Tiengo}, {Campana}, {Mereghetti}, {Rodriguez Castillo}, {G{\"o}tz},
  {Burgay}, {Possenti}, {Zane}, {Turolla}, {Perna}, {Cannizzaro}, \&
  {Pons}}]{israel16}
{Israel}, G.~L., {Esposito}, P., {Rea}, N., {et~al.} 2016, \mnras, 457, 3448

\bibitem[{{Kaspi} \& {Beloborodov}(2017)}]{kaspi17}
{Kaspi}, V.~M., \& {Beloborodov}, A.~M. 2017, \araa, 55, 261

\bibitem[{{Kennea} {et~al.}(2020){Kennea}, {Beardmore}, {Page}, \&
  {Palmer}}]{kennea20}
{Kennea}, J.~A., {Beardmore}, A.~P., {Page}, K.~L., \& {Palmer}, D.~M. 2020,
  The Astronomer's Telegram, 13679, 1

\bibitem[{{Li} {et~al.}(2020){Li}, {Lin}, {Xiong}, {Ge}, {Li}, {Li}, {Lu},
  {Zhang}, {Tuo}, {Nang}, {Zhang}, {Xiao}, {Chen}, {Song}, {Xu}, {Liu}, {Jia},
  {Cao}, {Zhang}, {Qu}, {Liao}, {Zhao}, {Tan}, {Nie}, {Zhao}, {Zheng}, {Zheng},
  {Luo}, {Cai}, {Li}, {Xue}, {Bu}, {Chang}, {Chen}, {Chen}, {Chen}, {Chen},
  {Chen}, {Cui}, {Cui}, {Deng}, {Dong}, {Du}, {Fu}, {Gao}, {Gao}, {Gao}, {Gu},
  {Guan}, {Guo}, {Han}, {Huang}, {Huo}, {Jiang}, {Jiang}, {Jin}, {Jin}, {Kong},
  {Li}, {Li}, {Li}, {Li}, {Li}, {Li}, {Li}, {Liang}, {Liu}, {Liu}, {Liu},
  {Liu}, {Liu}, {Lu}, {Lu}, {Luo}, {Ma}, {Meng}, {Ou}, {Sai}, {Shang}, {Song},
  {Sun}, {Tao}, {Wang}, {Wang}, {Wang}, {Wang}, {Wang}, {Wen}, {Wu}, {Wu},
  {Wu}, {Xiao}, {Yang}, {Yang}, {Yang}, {Yang}, {Yi}, {Yin}, {You}, {Zhang},
  {Zhang}, {Zhang}, {Zhang}, {Zhang}, {Zhang}, {Zhang}, {Zhang}, {Zhang},
  {Zhang}, {Zhang}, {Zhang}, {Zhang}, {Zhang}, {Zhang}, {Zhang}, {Zhou},
  {Zhou}, {Zhu}, {Zhu}, \& {Zhuang}}]{li20}
{Li}, C.~K., {Lin}, L., {Xiong}, S.~L., {et~al.} 2020, arXiv e-prints,
  arXiv:2005.11071

\bibitem[{{Lin} {et~al.}(2020{\natexlab{a}}){Lin}, {G{\"o}{\u g}{\"u}{\c s}},
  {Roberts}, {Kouveliotou}, {Kaneko}, {van der Horst}, \& {Younes}}]{lin20a}
{Lin}, L., {G{\"o}{\u g}{\"u}{\c s}}, E., {Roberts}, O.~J., {et~al.}
  2020{\natexlab{a}}, \apj, 893, 156

\bibitem[{{Lin} {et~al.}(2020{\natexlab{b}}){Lin}, {Zhang}, {Wang}, {Gao},
  {Guan}, {Han}, {Jiang}, {Jiang}, {Lee}, {Li}, {Men}, {Miao}, {Niu}, {Niu},
  {Sun}, {Wang}, {Wang}, {Xu}, {Xu}, {Xu}, {Yang}, {Yang}, {Yu}, {Zhang},
  {Zhang}, {Zhou}, {Zhu}, {Castro-Tirado}, {Dai}, {Ge}, {Hu}, {Li}, {Li}, {Li},
  {Liang}, {Jia}, {Querel}, {Shao}, {Wang}, {Wang}, {Wu}, {Xiong}, {Xu},
  {Yang}, {Zhang}, {Zhang}, {Zheng}, \& {Zou}}]{lin20b}
{Lin}, L., {Zhang}, C.~F., {Wang}, P., {et~al.} 2020{\natexlab{b}}, arXiv
  e-prints, arXiv:2005.11479

\bibitem[{{Margalit} {et~al.}(2020){Margalit}, {Beniamini}, {Sridhar}, \&
  {Metzger}}]{margalit20}
{Margalit}, B., {Beniamini}, P., {Sridhar}, N., \& {Metzger}, B.~D. 2020, arXiv
  e-prints, arXiv:2005.05283

\bibitem[{{Mereghetti} {et~al.}(2020){Mereghetti}, {Savchenko}, {Ferrigno},
  {G{\"o}tz}, {Rigoselli}, {Tiengo}, {Bazzano}, {Bozzo}, {Coleiro},
  {Courvoisier}, {Doyle}, {Goldwurm}, {Hanlon}, {Jourdain}, {von Kienlin},
  {Lutovinov}, {Martin-Carrillo}, {Molkov}, {Natalucci}, {Onori}, {Panessa},
  {Rodi}, {Rodriguez}, {S{\'a}nchez-Fern{\'a}ndez}, {Sunyaev}, \&
  {Ubertini}}]{mereghetti20}
{Mereghetti}, S., {Savchenko}, V., {Ferrigno}, C., {et~al.} 2020, arXiv
  e-prints, arXiv:2005.06335

\bibitem[{{Palmer} \& {BAT Team}(2020)}]{palmer20}
{Palmer}, D.~M., \& {BAT Team}. 2020, GRB Coordinates Network, 27665, 1

\bibitem[{{Perna} \& {Pons}(2011)}]{perna11}
{Perna}, R., \& {Pons}, J.~A. 2011, \apjl, 727, L51

\bibitem[{{Petroff} {et~al.}(2019){Petroff}, {Hessels}, \&
  {Lorimer}}]{petroff19}
{Petroff}, E., {Hessels}, J.~W.~T., \& {Lorimer}, D.~R. 2019, \aapr, 27, 4

\bibitem[{{Pons} \& {Rea}(2012)}]{pons12}
{Pons}, J.~A., \& {Rea}, N. 2012, \apjl, 750, L6

\bibitem[{{Popov} \& {Postnov}(2013)}]{popov13}
{Popov}, S.~B., \& {Postnov}, K.~A. 2013, arXiv e-prints, arXiv:1307.4924

\bibitem[{{Rea} {et~al.}(2016){Rea}, {Borghese}, {Esposito}, {Coti Zelati},
  {Bachetti}, {Israel}, \& {De Luca}}]{rea16}
{Rea}, N., {Borghese}, A., {Esposito}, P., {et~al.} 2016, \apjl, 828, L13

\bibitem[{{Rea} {et~al.}(2012{\natexlab{a}}){Rea}, {Pons}, {Torres}, \&
  {Turolla}}]{rea12b}
{Rea}, N., {Pons}, J.~A., {Torres}, D.~F., \& {Turolla}, R. 2012{\natexlab{a}},
  \apjl, 748, L12

\bibitem[{{Rea} {et~al.}(2010){Rea}, {Esposito}, {Turolla}, {Israel}, {Zane},
  {Stella}, {Mereghetti}, {Tiengo}, {G{\"o}tz}, {G{\"o}{\u g}{\"u}{\c s}}, \&
  {Kouveliotou}}]{rea10}
{Rea}, N., {Esposito}, P., {Turolla}, R., {et~al.} 2010, Science, 330, 944

\bibitem[{{Rea} {et~al.}(2012{\natexlab{b}}){Rea}, {Israel}, {Esposito},
  {Pons}, {Camero-Arranz}, {Mignani}, {Turolla}, {Zane}, {Burgay}, {Possenti},
  {Campana}, {Enoto}, {Gehrels}, {G{\"o}{\u g}{\"u}{\c s}}, {G{\"o}tz},
  {Kouveliotou}, {Makishima}, {Mereghetti}, {Oates}, {Palmer}, {Perna},
  {Stella}, \& {Tiengo}}]{rea12a}
{Rea}, N., {Israel}, G.~L., {Esposito}, P., {et~al.} 2012{\natexlab{b}}, \apj,
  754, 27

\bibitem[{{Spitler} {et~al.}(2014){Spitler}, {Cordes}, {Hessels}, {Lorimer},
  {McLaughlin}, {Chatterjee}, {Crawford}, {Deneva}, {Kaspi}, {Wharton},
  {Allen}, {Bogdanov}, {Brazier}, {Camilo}, {Freire}, {Jenet},
  {Karako-Argaman}, {Knispel}, {Lazarus}, {Lee}, {van Leeuwen}, {Lynch},
  {Ransom}, {Scholz}, {Siemens}, {Stairs}, {Stovall}, {Swiggum},
  {Venkataraman}, {Zhu}, {Aulbert}, \& {Fehrmann}}]{spitler14}
{Spitler}, L.~G., {Cordes}, J.~M., {Hessels}, J.~W.~T., {et~al.} 2014, \apj,
  790, 101

\bibitem[{{Spitler} {et~al.}(2016){Spitler}, {Scholz}, {Hessels}, {Bogdanov},
  {Brazier}, {Camilo}, {Chatterjee}, {Cordes}, {Crawford}, {Deneva}, {Ferdman},
  {Freire}, {Kaspi}, {Lazarus}, {Lynch}, {Madsen}, {McLaughlin}, {Patel},
  {Ransom}, {Seymour}, {Stairs}, {Stappers}, {van Leeuwen}, \&
  {Zhu}}]{spitler16}
{Spitler}, L.~G., {Scholz}, P., {Hessels}, J.~W.~T., {et~al.} 2016, \nat, 531,
  202

\bibitem[{{Stamatikos} {et~al.}(2014){Stamatikos}, {Malesani}, {Page}, \&
  {Sakamoto}}]{stamatikos14}
{Stamatikos}, M., {Malesani}, D., {Page}, K.~L., \& {Sakamoto}, T. 2014, GRB
  Coordinates Network, 16520, 1

\bibitem[{{Tavani} {et~al.}(2020){Tavani}, {Casentini}, {Ursi}, {Verrecchia},
  {Addis}, {Antonelli}, {Argan}, {Barbiellini}, {Baroncelli}, {Bernardi},
  {Bianchi}, {Bulgarelli}, {Caraveo}, {Cardillo}, {Cattaneo}, {Chen}, {Costa},
  {Del Monte}, {Di Cocco}, {Di Persio}, {Donnarumma}, {Evangelista}, {Feroci},
  {Ferrari}, {Fioretti}, {Fuschino}, {Galli}, {Gianotti}, {Giuliani},
  {Labanti}, {Lazzarotto}, {Lipari}, {Longo}, {Lucarelli}, {Magro},
  {Marisaldi}, {Mereghetti}, {Morelli}, {Morselli}, {Naldi}, {Pacciani},
  {Parmiggiani}, {Paoletti}, {Pellizzoni}, {Perri}, {Perotti}, {Piano},
  {Picozza}, {Pilia}, {Pittori}, {Puccetti}, {Pupillo}, {Rapisarda},
  {Rappoldi}, {Rubini}, {Setti}, {Soffitta}, {Trifoglio}, {Trois},
  {Vercellone}, {Vittorini}, {Giommi}, \& {D' Amico}}]{tavani20}
{Tavani}, M., {Casentini}, C., {Ursi}, A., {et~al.} 2020, arXiv e-prints,
  arXiv:2005.12164

\bibitem[{{The CHIME/FRB Collaboration} {et~al.}(2020){The CHIME/FRB
  Collaboration}, {:}, {Andersen}, {Band ura}, {Bhardwaj}, {Bij}, {Boyce},
  {Boyle}, {Brar}, {Cassanelli}, {Chawla}, {Chen}, {Cliche}, {Cook},
  {Cubranic}, {Curtin}, {Denman}, {Dobbs}, {Dong}, {Fandino}, {Fonseca},
  {Gaensler}, {Giri}, {Good}, {Halpern}, {Hill}, {Hinshaw}, {H{\"o}fer},
  {Josephy}, {Kania}, {Kaspi}, {Landecker}, {Leung}, {Li}, {Lin}, {Masui},
  {Mckinven}, {Mena-Parra}, {Merryfield}, {Meyers}, {Michilli}, {Milutinovic},
  {Mirhosseini}, {M{\"u}nchmeyer}, {Naidu}, {Newburgh}, {Ng}, {Patel}, {Pen},
  {Pinsonneault-Marotte}, {Pleunis}, {Quine}, {Rafiei-Ravandi}, {Rahman},
  {Ransom}, {Renard}, {Sanghavi}, {Scholz}, {Shaw}, {Shin}, {Siegel}, {Singh},
  {Smegal}, {Smith}, {Stairs}, {Tan}, {Tendulkar}, {Tretyakov}, {Vanderlinde},
  {Wang}, {Wulf}, \& {Zwaniga}}]{scholz20}
{The CHIME/FRB Collaboration}, {:}, {Andersen}, B.~C., {et~al.} 2020, arXiv
  e-prints, arXiv:2005.10324

\bibitem[{van~der Walt {et~al.}(2011)van~der Walt, Colbert, \&
  Varoquaux}]{vanderWalt11}
van~der Walt, S., Colbert, S.~C., \& Varoquaux, G. 2011, Comput. Sci. Eng., 13,
  22

\bibitem[{{Vigan{\`o}} {et~al.}(2013){Vigan{\`o}}, {Rea}, {Pons}, {Perna},
  {Aguilera}, \& {Miralles}}]{vigano13}
{Vigan{\`o}}, D., {Rea}, N., {Pons}, J.~A., {et~al.} 2013, \mnras, 434, 123

\bibitem[{{Wilms} {et~al.}(2000){Wilms}, {Allen}, \& {McCray}}]{wilms00}
{Wilms}, J., {Allen}, A., \& {McCray}, R. 2000, \apj, 542, 914

\bibitem[{{Younes} {et~al.}(2017{\natexlab{a}}){Younes}, {Baring},
  {Kouveliotou}, {Harding}, {Donovan}, {G{\"o}{\u g}{\"u}{\c s}}, {Kaspi}, \&
  {Granot}}]{ybk17}
{Younes}, G., {Baring}, M.~G., {Kouveliotou}, C., {et~al.} 2017{\natexlab{a}},
  \apj, 851, 17

\bibitem[{{Younes} {et~al.}(2017{\natexlab{b}}){Younes}, {Kouveliotou},
  {Jaodand}, {Baring}, {van der Horst}, {Harding}, {Hessels}, {Gehrels},
  {Gill}, {Huppenkothen}, {Granot}, {G{\"o}{\u g}{\"u}{\c s}}, \&
  {Lin}}]{younes17}
{Younes}, G., {Kouveliotou}, C., {Jaodand}, A., {et~al.} 2017{\natexlab{b}},
  \apj, 847, 85

\bibitem[{{Younes} {et~al.}(2020){Younes}, {Baring}, {Kouveliotou},
  {Arzoumanian}, {Enoto}, {Doty}, {Gendreau},
  {G{\"o}{\u{g}}{\"u}{\textcommabelow s}}, {Guillot}, {G{\"u}ver}, {Harding},
  {Ho}, {van der Horst}, {Jaisawal}, {Kaneko}, {LaMarr}, {Lin}, {Majid},
  {Okajima}, {Pope}, {Ray}, {Roberts}, {Saylor}, {Steiner}, \&
  {Wadiasingh}}]{younes20}
{Younes}, G., {Baring}, M.~G., {Kouveliotou}, C., {et~al.} 2020, arXiv
  e-prints, arXiv:2006.11358

\bibitem[{{Zhou} {et~al.}(2020){Zhou}, {Zhou}, {Chen}, {Wang}, {Vink}, \&
  {Wang}}]{zhou20}
{Zhou}, P., {Zhou}, X., {Chen}, Y., {et~al.} 2020, arXiv e-prints,
  arXiv:2005.03517

\end{thebibliography}

\end{document}